\def\fullversionflag{1}

\documentclass[conference]{IEEEtran}

\usepackage{etoolbox}

\newtoggle{fullversion}
\ifodd\fullversionflag
    \toggletrue{fullversion}
\else
    \togglefalse{fullversion}
\fi

\usepackage[hang,flushmargin]{footmisc}
\usepackage{hyperref}
\hypersetup{
  colorlinks=true,
  linkcolor={blue!60!black},
  citecolor={blue!60!black},
  urlcolor={blue!60!black},
}
\usepackage{microtype}
\usepackage{pdfsync}

\usepackage{amsthm}
\usepackage{amsmath}
\usepackage{amsfonts}
\usepackage{amssymb}
\usepackage{url}
\usepackage{graphicx}
\usepackage{mathtools}
\usepackage{footnote}
\usepackage{tabularx}
\usepackage{booktabs}
\usepackage{comment}
\usepackage{xspace}
\usepackage{multirow}
\usepackage{enumitem}
\usepackage{framed}
\usepackage{caption}
\usepackage{tikz}
\usepackage{breakcites}
\usepackage{subcaption}
\usepackage[capitalize]{cleveref}
\usepackage{stmaryrd}

\numberwithin{equation}{section}

\usepackage{tikz}
\usetikzlibrary{calc,positioning,shapes,shadows,arrows,fit}
\usepackage{pgfplots}
\usepackage{xcolor}
\usepackage{colortbl}

\newcommand{\ignore}[1]{}

\theoremstyle{plain} \newtheorem{theorem}{Theorem}[section]

\theoremstyle{definition} 
\newtheorem{definition}[theorem]{Definition}  

\newtheorem{remark}[theorem]{Remark}

\newcommand{\N}{\mathbb{N}}
\newcommand{\Z}{\mathbb{Z}}
\newcommand{\R}{\mathbb{R}}

\newcommand{\ms}[1]{\mathsf{#1}}

\newcommand{\getsr}{\xleftarrow{\textsc{r}}}
\newcommand{\abs}[1]{| #1 |}
\newcommand{\set}[1]{\{ #1 \}}

\newcommand{\etal}{et~al.\xspace}

\newcommand{\calD}{\ensuremath{\mathcal{D}}}

\newcommand{\calS}{\ensuremath{\mathcal{S}}}

\newcommand{\boldx}{\ensuremath{\mathbf{x}}}
\newcommand{\boldy}{\ensuremath{\mathbf{y}}}
\newcommand{\boldz}{\ensuremath{\mathbf{z}}}

\newcommand{\xv}{\boldx}
\newcommand{\yv}{\boldy}
\newcommand{\zv}{\boldz}

\newcommand{\boldA}{\ensuremath{\mathbf{A}}}
\newcommand{\boldB}{\ensuremath{\mathbf{B}}}

\newcommand{\Am}{\boldA}
\newcommand{\Bm}{\boldB}

\newcommand{\appc}{\stackrel{c}{\approx}}

\newcommand{\deq}{\mathrel{\mathop:}=}
\newcommand{\zo}{\ensuremath{\{0,1\}}} %

\renewcommand{\paragraph}[1]{\medskip \noindent \textbf{#1}}

\newcommand{\ring}{\Z_n}
\newcommand{\share}[1]{\llbracket #1 \rrbracket^n}
\newcommand{\bshare}[1]{\llbracket #1 \rrbracket^2}
\newcommand{\ord}[1]{#1^{\text{th}}}

\newcommand{\aby}{\ensuremath{\text{ABY}^3}\xspace}

\newcommand{\view}{\ms{view}}
\newcommand{\out}{\ms{output}}
\newcommand{\relu}{\ms{ReLU}}

\newcommand{\oursystem}{\ensuremath{\textsc{CryptGPU}}\xspace}

\newcommand{\Share}{\ms{Share}}
\newcommand{\Reconstruct}{\ms{Reconstruct}}

\newcommand{\pytorch}{\text{PyTorch}\xspace}
\newcommand{\crypten}{\textsc{CrypTen}\xspace}

\newcommand{\mpctensor}{\texttt{MPCTensor}\xspace}
\newcommand{\cudalong}{\texttt{CUDALongTensor}\xspace}

\newcommand{\softmax}{\mathrm{softmax}}

\pgfplotsset{%
  plot1/.style = {%
    green!60!black,
    mark=*,
    ultra thick
  },
  plot2/.style = {%
    purple!50!blue,
    mark=square*,
    ultra thick
  }
}

\usepackage{authblk}

\begin{document}

\title{\LARGE \oursystem: Fast Privacy-Preserving Machine Learning on the GPU}

\author{Sijun Tan\IEEEauthorrefmark{1},
        Brian Knott\IEEEauthorrefmark{2},
        Yuan Tian\IEEEauthorrefmark{1}, and
        David J. Wu\IEEEauthorrefmark{1} \vspace{0.5em} \\ 
  \IEEEauthorrefmark{1}University of Virginia \vspace{-0.2em} \\
  {\small \{\href{mailto:st8eu@virginia.edu}{\texttt{st8eu}}, 
            \href{mailto:yuant@virginia.edu}{\texttt{yuant}},
            \href{mailto:dwu4@virginia.edu}{\texttt{dwu4}}\}\texttt{@virginia.edu}} \vspace{0.2em} \\
  \IEEEauthorrefmark{2}Facebook AI Research \vspace{-0.2em} \\
  {\small \href{brianknott@fb.com}{\texttt{brianknott@fb.com}}}}

\maketitle

\pagestyle{plain}


\pgfplotstableread{
iter  priv     plain
0     2.303020 2.302111
20    2.309026 2.301171
40    2.307021 2.301428
60    2.305281 2.299104
80    2.300695 2.295183
100   2.297590 2.291928
120   2.299509 2.288321
140   2.286862 2.283239
160   2.284751 2.283605
180   2.276913 2.278234
200   2.271689 2.272111
220   2.244105 2.265531
240   2.226574 2.256311
260   2.197375 2.241122
280   2.148133 2.212307
300   2.080836 2.185154
320   1.984310 2.134359
340   1.865909 2.066423
360   1.841929 1.975379
380   1.806365 1.913700
400   1.779879 1.823409
420   1.754965 1.785549
440   1.662993 1.742904
460   1.642900 1.738635
480   1.651880 1.677799
500   1.606548 1.689598
520   1.550554 1.646766
540   1.638639 1.625648
560   1.493766 1.676247
580   1.528061 1.674721
600   1.522800 1.563667
620   1.474721 1.557829
640   1.454481 1.571985
660   1.404543 1.548808
680   1.468899 1.533960
700   1.442692 1.599770
720   1.382357 1.494472
740   1.365128 1.434306
760   1.360383 1.358962
780   1.394712 1.396999
800   1.346911 1.466320
820   1.360089 1.349542
840   1.353446 1.473161
860   1.302167 1.317898
880   1.231720 1.364884
900   1.349070 1.383716
920   1.250846 1.300275
940   1.334885 1.353355
960   1.229935 1.360966
980   1.241312 1.338423
1000  1.356414 1.290717
1020  1.301565 1.324881
1040  1.210126 1.364201
1060  1.270226 1.245869
1080  1.182453 1.355423
1100  1.246361 1.252801
1120  1.226410 1.253057
1140  1.322738 1.248731
1160  1.267931 1.264047
1180  1.201581 1.197805
1200  1.206633 1.307972
1220  1.190294 1.320832
1240  1.183639 1.161867
1260  1.180487 1.245801
1280  1.264293 1.112405
1300  1.271656 1.286762
1320  1.163678 1.159915
1340  1.136541 1.224848
1360  1.160552 1.218196
1380  1.162467 1.205675
1400  1.110052 1.201814
1420  1.066695 1.194778
1440  1.125360 1.227322
1460  1.207083 1.058508
1480  1.156292 1.173757
1500  1.103554 1.204694
1520  1.095647 1.124189
1540  1.128274 1.084311
1562  1.092242 1.066893
}\cifartraining

\pgfplotstableread{
iter  priv     plain
0     2.307991 2.297481
20    2.304085 2.285256
40    2.296361 2.272705
60    2.288371 2.256771
80    2.280919 2.236051
100   2.267573 2.216968
120   2.253341 2.179103
140   2.235681 2.137586
160   2.207279 2.063185
180   2.176018 1.941974
200   2.124168 1.808315
220   2.057989 1.603645
240   1.953015 1.322086
260   1.826064 1.093246
280   1.608181 0.918154
300   1.367639 0.809896
320   1.133011 0.700017
340   0.950375 0.657067
360   0.835743 0.658232
380   0.741961 0.617983
400   0.690795 0.588218
420   0.642913 0.560717
440   0.595956 0.539831
460   0.573195 0.522965
480   0.570364 0.510984
500   0.522368 0.486590
520   0.516423 0.483419
540   0.537180 0.452036
560   0.515444 0.525710
580   0.483199 0.502488
600   0.483147 0.454981
620   0.449009 0.449929
640   0.445914 0.437142
660   0.473097 0.429995
680   0.463074 0.430284
700   0.452394 0.435522
720   0.421640 0.418436
740   0.419794 0.365272
760   0.403595 0.396429
780   0.404266 0.410526
800   0.397624 0.387778
820   0.392634 0.389539
840   0.404237 0.356338
860   0.383337 0.370809
880   0.409347 0.357042
900   0.383942 0.410707
920   0.366925 0.373362
940   0.338403 0.370784
960   0.367384 0.371342
980   0.367823 0.329439
1000  0.378333 0.345761
1020  0.345816 0.355993
1040  0.343119 0.382331
1060  0.326913 0.345660
1080  0.347786 0.341994
1100  0.358791 0.351213
1120  0.351472 0.328825
1140  0.345753 0.305150
1160  0.333702 0.330232
1180  0.326428 0.344555
1200  0.333553 0.325817
1220  0.317538 0.320010
1240  0.330949 0.299318
1260  0.330888 0.314067
1280  0.306789 0.331741
1300  0.338620 0.320798
1320  0.310727 0.298246
1340  0.310625 0.322673
1360  0.311785 0.284305
1380  0.323229 0.275806
1400  0.305269 0.306588
1420  0.323037 0.293995
1440  0.317729 0.305910
1460  0.296127 0.271164
1480  0.287875 0.274589
1500  0.259893 0.278428
1520  0.314888 0.287768
1540  0.292759 0.277439
1560  0.276390 0.257709
1580  0.275094 0.286122
1600  0.274542 0.294305
1620  0.275453 0.262447
1640  0.301770 0.252409
1660  0.231212 0.279187
1680  0.267743 0.267802
1700  0.279838 0.263005
1720  0.280268 0.286263
1740  0.272248 0.271808
1760  0.226357 0.256103
1780  0.259859 0.248243
1800  0.279499 0.279655
1820  0.270261 0.248574
1840  0.264414 0.283972
1860  0.254663 0.238629
1880  0.263624 0.260379
1900  0.232964 0.254015
1920  0.251797 0.222183
1940  0.234401 0.235458
1960  0.255717 0.245033
1980  0.247042 0.223650
2000  0.264214 0.256884
2020  0.239969 0.239208
2040  0.239974 0.228044
2060  0.253987 0.234931
2080  0.252018 0.208952
2100  0.248576 0.237355
2120  0.210640 0.226911
2140  0.250735 0.232681
2160  0.209687 0.248925
2180  0.217858 0.223947
2200  0.218142 0.232719
2220  0.223970 0.241114
2240  0.234176 0.202158
2260  0.198709 0.219779
2280  0.217929 0.246622
2300  0.203081 0.212930
2320  0.203704 0.207259
2344  0.204614 0.212115
}\mnisttraining

\pgfplotstableread{
iter  priv     plain   
0     5.348772 5.292922
20    5.351716 5.297915
40    5.358248 5.296045
60    5.354621 5.296695
80    5.347533 5.295563
100   5.356356 5.293538
120   5.354140 5.299761
140   5.354647 5.299225
160   5.351066 5.294886
180   5.349693 5.295034
200   5.357671 5.297765
220   5.348639 5.294207
240   5.348551 5.294084
260   5.350354 5.291862
280   5.346176 5.288854
300   5.344809 5.292825
320   5.349773 5.294276
340   5.346973 5.296135
360   5.347289 5.290172
380   5.347751 5.290251
400   5.346366 5.292538
420   5.351233 5.283771
440   5.350850 5.293459
460   5.346487 5.287403
480   5.348024 5.285364
500   5.343609 5.288424
520   5.344928 5.284933
540   5.344568 5.283842
560   5.342005 5.285740
580   5.340844 5.283682
600   5.342674 5.282900
620   5.339786 5.280702
640   5.340684 5.280615
660   5.340478 5.278415
680   5.335614 5.276947
700   5.337988 5.277703
720   5.331269 5.270405
740   5.338122 5.275547
760   5.332749 5.262685
780   5.335867 5.269873
800   5.334847 5.263180
820   5.329391 5.262619
840   5.324759 5.248409
860   5.330606 5.263144
880   5.323787 5.251213
900   5.320579 5.244101
920   5.314856 5.226041
940   5.306584 5.227245
960   5.310904 5.229631
980   5.312452 5.205789
1000  5.297267 5.188951
1020  5.292648 5.183874
1040  5.281113 5.155966
1060  5.282702 5.156560
1080  5.266032 5.104183
1100  5.233905 5.077445
1120  5.222896 5.064382
1140  5.204712 5.038204
1160  5.198780 5.011984
1180  5.208151 4.997435
1200  5.171941 4.964686
1220  5.176840 4.926416
1240  5.133299 4.897495
1260  5.101889 4.834286
1280  5.039693 4.841530
1300  5.054780 4.764454
1320  5.033660 4.744911
1340  4.980012 4.751329
1360  4.974542 4.626973
1380  4.894507 4.621947
1400  4.946734 4.633847
1420  4.877776 4.628491
1440  4.846462 4.527771
1460  4.826614 4.466598
1480  4.793755 4.556854
1500  4.687559 4.530313
1520  4.775496 4.507755
1540  4.624485 4.490775
1560  4.691902 4.434336
1580  4.665820 4.475128
1600  4.582723 4.410460
1620  4.650419 4.317559
1640  4.525076 4.340008
1660  4.588584 4.301797
1680  4.614716 4.313292
1700  4.501956 4.200781
1720  4.485007 4.267214
1740  4.486487 4.308079
1760  4.403178 4.216676
1780  4.450204 4.175800
1800  4.420725 4.132192
1820  4.333404 4.154937
1840  4.321035 4.161268
1860  4.370826 4.269868
1880  4.325295 4.127577
1900  4.243224 4.054141
1920  4.276851 4.097298
1940  4.201192 4.117095
1960  4.281686 4.081095
1980  4.218047 3.990636
2000  4.150362 4.055569
2020  4.173797 3.981065
2040  4.196634 4.092000
2060  4.142936 4.135289
2080  4.147063 3.997328
2100  4.068864 4.079680
2120  4.144876 4.060939
2140  4.163410 4.049058
2160  4.141045 3.908458
2180  4.055923 4.028896
2200  4.155239 3.977061
2220  4.103526 3.933063
2240  4.098282 3.950396
2260  4.097110 3.920829
2280  4.012187 3.902432
2300  4.039274 3.971662
2320  4.025286 3.892863
2340  4.094119 3.926284
2360  3.934062 3.877881
2380  4.005063 3.852662
2400  3.975772 3.921389
2420  4.031226 3.825790
2440  3.955954 3.887264
2460  3.981359 3.857487
2480  3.902600 3.917298
2500  3.909162 3.826680
2520  3.839953 3.875010
2540  3.949748 3.841827
2560  3.844684 3.836207
2580  3.817641 3.852584
2600  3.854777 3.802909
2620  3.830697 3.838158
2640  3.885766 3.800232
2660  3.850523 3.800855
2680  3.984241 3.793755
2700  3.898606 3.707421
2720  3.798808 3.662229
2740  3.820111 3.708503
2760  3.758666 3.653278
2780  3.778158 3.705590
2800  3.748090 3.690548
2820  3.855549 3.760042
2840  3.804991 3.661139
2860  3.744910 3.650055
2880  3.876588 3.713900
2900  3.799416 3.762305
2920  3.808065 3.679557
2940  3.764808 3.735645
2960  3.816901 3.728475
2980  3.773047 3.678527
3000  3.802283 3.740670
3020  3.719207 3.600734
3040  3.877991 3.627173
3060  3.658016 3.596598
3080  3.743847 3.654312
3100  3.712750 3.622238
3124  3.765742 3.588078
}\titraining


\begin{abstract}
  We introduce \oursystem, a system for privacy-preserving machine learning that
  implements {\em all} operations 
  on the GPU (graphics processing unit). Just as GPUs played a pivotal role in the
  success of modern deep learning, they are also essential for realizing scalable
  {\em privacy-preserving} deep learning. In this work, we start by
  introducing a new interface to losslessly embed cryptographic operations over secret-shared
  values (in a {\em discrete} domain) into floating-point operations that can be processed
  by highly-optimized CUDA kernels for linear algebra. We then identify a sequence of
  ``GPU-friendly'' cryptographic protocols to enable privacy-preserving evaluation of
  both linear {\em and} non-linear operations on the GPU. Our microbenchmarks indicate that our private
  GPU-based convolution protocol is over $150\times$ faster than the analogous CPU-based protocol;
  for non-linear operations like the ReLU activation function, our GPU-based protocol is around $10\times$
  faster than its CPU analog.

  With \oursystem, we support private inference {\em and} private training on convolutional neural networks with
  over 60 million parameters as well as handle large datasets like
  ImageNet. Compared to the previous state-of-the-art,
  when considering large models and datasets,
  our protocols achieve a $2\times$ to $8\times$
  improvement in private inference and a $6\times$ to $36\times$
  improvement for private training. Our
  work not only showcases the viability of performing secure multiparty computation (MPC)
  {\em entirely} on the GPU to enable fast privacy-preserving
  machine learning, but also highlights the importance of designing new MPC primitives that can
  take full advantage of the GPU's computing capabilities.
\end{abstract}


\section{Introduction}

Deep learning has enabled numerous
applications in the form of digital voice assistants, video
monitoring and surveillance systems, and even systems for disease diagnosis
and treatment planning. But these new and exciting applications raise
challenging questions regarding user privacy. After all, modern machine
learning algorithms are largely data-driven and training image recognition,
speech recognition, or disease predictor systems all rely on aggregating and
analyzing sensitive user data. Even model inference raises privacy
concerns as increasingly often, voice or video recordings from a mobile or IoT
device are outsourced to the cloud for analysis.

To address some of the privacy challenges associated with the widespread deployment
of deep learning technologies,
a number of works~\cite{MZ17,MR18,WGC19,MLSZP20,KRCGRS20,WTBKMR21} in the last few years have
introduced cryptographic frameworks based on secure multiparty computation (MPC)~\cite{GMW87,BGW88}
to enable {\em privacy-preserving deep learning}
(see \cref{sec:related} for a more comprehensive
survey). At a high level, MPC protocols
allow a set of mutually-distrusting parties to compute an arbitrary function over secret
inputs such that at the end of the computation, the parties only learn the output of
their computation, and nothing more. In particular, all information about other parties'
inputs are completely hidden (up to what could be inferred based on the output\footnote{There
are settings where even learning the exact output is problematic and can reveal
compromising information about other parties' inputs. Techniques like
differential privacy~\cite{SS15,ACGMMTZ16} provide a defense against these types of attacks. We discuss this
in greater detail in \cref{sec:related}.}).

While there have been considerable advances in the concrete efficiency of MPC protocols,
current approaches remain computationally expensive and do not scale well to the types of
neural networks typically used in modern machine learning systems. Until recently,
cryptographic protocols for private inference over deep neural networks have been limited to
small datasets such as MNIST~\cite{MNIST} or CIFAR~\cite{CIFAR}.
In contrast, the current baseline for object recognition is ImageNet~\cite{ImageNet},
a dataset that is over $1000\times$ larger than CIFAR/MNIST and contains 1000 different classes (compared to just 10 classes for MNIST and CIFAR).
Similarly, state-of-the-art deep learning models for computer vision
such as ResNet-152~\cite{HZRS16} contain over 150 layers and over 60 million parameters.
In contrast, most protocols for privacy-preserving machine learning have been constrained to
relatively shallow networks with just tens of layers and a few hundred thousand parameters.

Recently, two systems \textsc{Falcon}~\cite{WTBKMR21} and
\textsc{CrypTFlow}~\cite{KRCGRS20} have made considerable headway towards scalable
privacy-preserving machine learning. For the first time, they demonstrate the ability
to perform privacy-preserving machine learning at the scale of ImageNet (or Tiny ImageNet~\cite{TinyImageNet}
in the case of \textsc{Falcon}) and with much larger models (e.g., AlexNet~\cite{KSH12}, VGG-16~\cite{SZ14},
and the ResNet family of models~\cite{HZRS16}). In spite of these
advances, there still remains considerable overhead: for example, private training of
AlexNet on Tiny ImageNet is estimated to still take {\em over a year}
using \textsc{Falcon}.
\textsc{CrypTFlow} currently only supports private inference and not private training. Both works
argue that hardware acceleration with graphics processing units (GPUs) will be essential
for scaling up privacy-preserving deep learning, especially in the case of private training.

\paragraph{The importance of GPU acceleration.} GPUs and hardware acceleration have played
a critical role in the evolution of modern deep learning. Today,
convolutional neural networks (CNNs) have become
a staple for modern computer vision. However, 
in the immediate years following their introduction in the seminal work of
LeCun~\etal~\cite{LBDHHHJ89},
CNNs did not see widespread adoption. This was in large
part due to the high computational costs of the
backpropagation training algorithm. Starting the mid-2000s,
several works~\cite{CPS06,CMGS10} showed that CNN training could be greatly accelerated through
the use of graphics processing units (GPUs). This culminated with the breakthrough moment when
Krizhevsky~\etal~\cite{KSH12} introduced ``AlexNet'' and 
won the ImageNet Large Scale Visual Recognition Challenge in 2012 using
a large CNN {\em trained entirely} on the GPU. Since AlexNet, CNNs have become a mainstay
of computer vision. Modern machine learning frameworks like \pytorch~\cite{PyTorch19}
and TensorFlow~\cite{TensorFlow16} all support and rely heavily on not only
GPUs, but even custom-designed
application-specific integrated circuits (ASICs) such as Google's tensor processing unit~\cite{TPU}.

\paragraph{Privacy-preserving machine learning on the GPU.}
Hardware acceleration has become a core component for evaluating and training
deep learning models. Given that MPC protocols necessarily incur a non-zero
overhead on top of the plaintext computation, it is essential for cryptographic
protocols to be able to leverage GPU acceleration in order to have {\em any} chance of scaling
up to support training and inference over deep models. After all, if we are bound to CPU-based
computations (as nearly {\em all} existing MPC frameworks have), then it is infeasible to even
run the machine learning algorithm on {\em plaintext} data. 

\subsection{Our Contributions}
\label{sec:our-contributions}

In this work, we introduce
\oursystem, a new cryptographic MPC framework built on top of \pytorch and \crypten~\cite{KVHSIM20}
where {\em all} of the cryptographic operations (both linear {\em and} non-linear) are implemented
on the GPU. \oursystem operates in the standard 3-party setting where we assume that all inputs are
secret-shared across
three non-colluding servers who execute the MPC protocol.
The inputs are secret shared using a 
2-out-of-3 {\em replicated secret sharing} scheme~\cite{ISN89,AFLNO16} (see \cref{sec:crypto} for the
full details).
Our system provides security against a single semi-honest
corruption. We describe our threat model formally in \cref{sec:threat-model}.

\oursystem can perform private inference over modern computer vision models such as
ResNet-152 on ImageNet images in just over 25s ($2.3\times$ faster than the previous state-of-the-art
\textsc{CrypTFlow}~\cite{KRCGRS20}).
For smaller networks like AlexNet, private
inference over ImageNet requires just 1.5s.

Further improvements to the costs of private training are possible
if we consider {\em batch inference}, which
also benefits from GPU parallelism. For example, batch inference over ResNet-152 reduces
the cost of private inference from 25s for a single image
to 13.2s per image when amortized over a batch of $8$ images.

For private training (which has a greater potential to benefit from GPU acceleration),
we demonstrate a $36\times$ speed-up for private training of AlexNet on the Tiny
ImageNet database compared to \textsc{Falcon}. Whereas it would have taken over a year to privately train
\textsc{Falcon} on Tiny ImageNet, our GPU-accelerated system would be
able to do so in just over a week (see \cref{sec:benchmarks-main}).
Beyond these performance results, our work highlights the potential of leveraging 
GPUs to accelerate privacy-preserving deep learning in much the same way GPUs have dramatically accelerated
standard deep learning. Our work also highlights the importance of developing new types of cryptographic 
protocols that are ``GPU-friendly'' and can take advantage of the parallelism provided by GPUs.

\paragraph{Cryptography on the GPU.} While NVIDIA's CUDA (Compute Unified Device Architecture) platform~\cite{CUDA}
supports general-purpose computations on the GPU, directly translating code written for the CPU
onto the GPU is unlikely to translate to immediate performance gains.
The architectural differences between
the CPU and the GPU introduce several additional hurdles that must be overcome
in order to have an efficient implementation:
\begin{itemize}[leftmargin=*]
  \item \textbf{Leveraging existing CUDA kernels.}
  The first challenge is that highly optimized CUDA kernels for computing deep learning primitives
  (i.e., convolutions, pooling, matrix multiplication) are designed to operate on {\em floating-point}
  inputs, and there does not currently exist kernels for computing on integer values. In MPC,
  we typically compute over discrete objects (i.e., ring or field elements). To leverage
  optimized kernels for these basic primitives, we need a way to {\em embed} the integer-valued
  cryptographic operations into (64-bit) floating-point arithmetic that can in turn be operated on by these
  kernels. \oursystem enables this by introducing
  a new abstraction called a \cudalong that models tensors (i.e., multi-dimensional
  arrays) of integer values, but seamlessly translates the integer-valued computations
  into a corresponding set of floating-point computations. We describe our construction in \cref{sec:sys-design}.
  The \textsc{Delphi} system encountered a similar challenge, but as we discuss
  in \cref{rmk:delphi-float}, their solution does not extend well to our setting. A critical difference
  is that \textsc{Delphi} considers private inference where
  the model is {\em public} while in this work, we assume that the model is also hidden (i.e., secret-shared).

  \item \textbf{``GPU-friendly'' cryptography.} The GPU architecture is optimized for performing a large number
  of {\em simple} computations on {\em blocks} of values. This means that operations like component-wise addition
  and multiplication of vectors/matrices are fast while operations that involve large numbers of conditional
  statements are slower. While there is support for integer addition and multiplication, operations like
  computing a modular reduction by a prime incurs considerably more overhead~\cite{CUDA}; for instance, we observed
  a $40\times$ difference in the running time of point-wise addition vs.
  point-wise modular reduction. Thus, when choosing
  and designing cryptographic protocols for the GPU, one must carefully calibrate them for the architecture.
  Protocols like Yao's garbled circuits~\cite{Yao86}
  are less well-suited
  for taking advantage of GPU parallelism compared to a vectorized secret-sharing-based protocol. Similarly,
  protocols that require extensive finite field arithmetic (and thus, require modular reductions) will incur
  more overhead on the GPU compared to protocols that only rely on arithmetic modulo a power of $2$.
  We also design protocols for common non-linear functions (e.g., exponentiation and division)
  that are specifically optimized for our particular setting.
  We describe the cryptographic protocols we use in~\cref{sec:crypto}.
\end{itemize}

\paragraph{Systematic evaluation of GPU-based MPC.} We present a comprehensive and systematic
evaluation of \oursystem to quantify the advantages of a GPU-based MPC protocol and compare against
previous protocols for privacy-preserving machine learning. We specifically measure the performance
of our private training and inference 
protocols on a wide range of object recognition models (e.g., LeNet~\cite{LBDHHHJ89}, AlexNet~\cite{KSH12},
and the ResNet family of networks~\cite{HZRS16}) and datasets (e.g., MNIST~\cite{MNIST}, CIFAR-10~\cite{CIFAR},
and ImageNet~\cite{ImageNet}). We describe our experimental methodology and measurements
in \cref{sec:exp}.

We also collect fine-grained measurements to
understand how the computational costs are split across the different layers of
a network. For instance, in CPU-based systems like \textsc{Falcon}~\cite{WTBKMR21},
the linear layers account for
$86$\% to $99$\% of the overall computational costs of private training.\footnote{While
linear layers are simpler to evaluate from a cryptographic perspective (in comparison
to non-linear layers), the size of the linear
layers is typically much larger than that of the non-linear layers.}
On the same model/datasets,
our GPU-based approach evaluates the same linear layers with a
$25\times$ to $72\times$ speed-up; this is a major source of the performance advantage
of \oursystem compared to previous systems. Consequently, the costs of
our private training protocol is more evenly split between evaluating linear layers
and non-linear layers. We provide the full details in \cref{sec:benchmarks-main,tab:falcon-comp-fine}.

In \cref{sec:micro}, we report microbenchmarks to quantify the performance
advantages of using the GPU to execute all of the MPC protocols. For instance, we show that
evaluating convolutions on secret-shared data (with secret-shared kernels) on the GPU is
over $150\times$ faster than the corresponding protocol on the CPU. 
Even for non-linear operations like the ReLU
(rectified linear unit) function, using a GPU-based MPC protocol still yields a
$10\times$ speed-up over the same underlying CPU-based protocol.

Finally, since our MPC protocol represents real-valued inputs using a fixed-point encoding,
and moreover, some of our protocols rely on approximations to non-linear functions, we also
compare the accuracy of our private inference and private training algorithms to
the analogous plaintext algorithms. As we show in \cref{sec:acc}, for the models and
datasets we consider in this work, the behavior of our privacy-preserving algorithms closely
matches their plaintext analogs.

\paragraph{An ML-friendly approach.} One of the guiding principles behind our system design
is to make it friendly for machine learning researchers to use. We build our system on top of 
\crypten~\cite{KVHSIM20}, which is itself built on top of the popular machine learning framework
PyTorch~\cite{PyTorch19}. Effectively, our work (much like \crypten) provides a new cryptographic 
{\em back end} that supports computations on secret-shared values while retaining a similar front end
as PyTorch. In fact, we note that our work on developing the \cudalong module has already been integrated
as part of \crypten to support privacy-preserving GPU computations~\cite{KVHSIM20}.


\section{System Overview}

Similar to previous works on constructing efficient protocols for
privacy-preserving machine learning~\cite{MR18,WGC19,WTBKMR21,KRCGRS20,CCPS19,PS20}
(see also \cref{sec:related}), we assume that
the data and model are (arbitrarily) partitioned across three parties. For
example, the three parties could be three independent organizations seeking to
collaboratively train a model on their joint data without revealing their
inputs to each other. Our system is also applicable in the ``server-aided''
setting~\cite{KMR11}, where a group of (arbitrarily-many) clients seek to train
a joint model on their data (or evaluate a secret-shared model on private
inputs). In the server-aided setting, the clients first secret share
their inputs to three independent cloud-service providers, who in turn run the
cryptographic protocol on the secret-shared inputs. We design our protocols
to provide security against a single semi-honest corruption. We provide a formal
description of our threat model in \cref{sec:threat-model}.

\subsection{Background}
\label{sec:architecture-background}

Our starting point in this work is the \crypten privacy-preserving machine
learning framework~\cite{KVHSIM20}. \crypten is built on top of the widely-used
machine-learning framework \pytorch~\cite{PyTorch19}. We adapt the basic architecture of \crypten, and make modifications to support three-party protocols based on replicated secret sharing. We describe the main system architecture below.

\paragraph{GPU architecture.} GPUs, and more recently,
ASICs like Google's tensor processing units~\cite{TPU},
have played a critical role in scaling up modern deep learning. These specialized
hardware platforms support massive parallelism, making them well-suited for performing
standard linear algebraic operations (e.g., convolutions or average pooling)
as well as point-wise evaluation of functions on large blocks of neurons (e.g., evaluating
an activation function or performing batch normalization).
Popular frameworks for deep learning frameworks such as \pytorch~\cite{PyTorch19}
and TensorFlow~\cite{TensorFlow16} natively support computations on both the CPU and GPU.

CUDA is a parallel computing platform
developed by NVIDIA for general-purpose computing on GPUs~\cite{CUDA}.
For deep learning in particular,
CUDA libraries such as \texttt{cuBLAS}~\cite{cuBLAS} and \texttt{cuDNN}~\cite{cuDNN}
provide highly-optimized implementation for a wide-range of standard primitives
such as convolutions, pooling, activation functions, and more. These libraries
are designed for {\em floating-point} computations and do {\em not} support
integer-valued analogs of these operations. Since cryptographic protocols typically
operate over {\em discrete} spaces (e.g., a 64-bit ring) where the underlying
algebra is implemented using integer-valued computations, one cannot directly
translate an existing protocol to the GPU.

\paragraph{\pytorch.} \pytorch~\cite{PyTorch19} is a popular open-source
machine learning framework designed for prototyping, implementing, and
deploying deep neural networks. The \pytorch front end supports many
standard neural network layers (e.g., convolutions, pooling, activation
functions, etc.) as well as features such as automatic differentiation and
gradient computation. The \pytorch back end natively supports
computation on both CPUs as well as GPUs. This flexibility enables
users to train complex models without needing to worry about the finer details
of backpropagation. It also allows users to take advantage of GPU acceleration without needing to
interface with low-level CUDA kernels. \pytorch also provides
library support for distributing computations across multiple devices and/or GPUs.

Data in \pytorch is organized around
{\em tensors}, which provide a general abstraction for
$n$-dimensional arrays. \pytorch provides an expressive API for computing on and applying
transformations to tensors. Especially importantly in our case, the \pytorch back end
natively and seamlessly leverages GPU acceleration for tensor computations. 

\paragraph{\crypten.} \crypten~\cite{KVHSIM20} is a recent framework built
on top of \pytorch for privacy-preserving machine learning.
\crypten provide a secure computing {\em back end} for \pytorch while still preserving
the \pytorch front end APIs that enables rapid prototyping and experimentation
with deep neural networks.

The main data abstraction in \crypten is the \mpctensor, which functions
like a standard \pytorch tensor, except the values are {\em secret shared} across
multiple machines. Internally, \crypten uses $n$-out-of-$n$ additive secret sharing.
For bilinear operations such as convolutions and matrix multiplications, \crypten
uses arithmetic secret sharing over a large ring (e.g., $\Z_{2^{64}}$), while for evaluating
non-linear operations like an activation function, it uses Boolean secret sharing.
\crypten uses the ABY share-conversion techniques~\cite{DSZ15} to convert between
arithmetic shares and Boolean shares.

\crypten supports general $n$-party computation and provides security against a single
semi-honest corruption. At the cryptographic level, elementary arithmetic
operations are handled using Beaver multiplication triples~\cite{Bea91},
Boolean circuit evaluation is implemented using the Goldreich-Micali-Wigderson
(GMW) protocol~\cite{GMW87}, and low-degree polynomial approximations are used
for most non-linear operations. We note that while our system builds on \crypten,
we work in a 3-party model where parties compute using {\em replicated
secret shares} (as in~\cite{AFLNO16}). We describe this in \cref{sec:crypto}.

\subsection{System Design and Architecture}
\label{sec:sys-design}

The design of \oursystem is centered around the following principles:
\begin{itemize}[leftmargin=*]
  \item \textbf{Leverage existing CUDA kernels for linear algebra.} As mentioned in
  \cref{sec:architecture-background}, highly-optimized CUDA kernels exist for most linear
  algebra operations encountered in deep learning. However, these kernels only support computations
  on floating-point values and are not directly applicable for computing on discrete structures
  common in cryptographic protocols. Thus, we seek a way to keep all of the
  computation on the GPU itself.

  \item \textbf{Keep {\em all} computations on the GPU.} While some previous works
  on private machine learning~\cite{TB19,MLSZP20} show how to leverage the GPU for computing
  linear and bilinear functions, they then move the data out of the GPU to evaluate non-linear functions.
  In this work, we seek to keep {\em all} of the computations on the GPU, and as we show in \cref{sec:micro},
  even computing non-linear functions can benefit greatly from GPU acceleration, provided that
  they are implemented using ``GPU-friendly'' cryptographic protocols (i.e., protocols that primarily
  rely on point-wise or component-wise vector operations).
\end{itemize}

\paragraph{Floating point computations.} The cryptographic core of \oursystem
relies on (additive) replicated secret sharing over the 64-bit ring $\Z_{2^{64}}$.
Computing bilinear functions such
as convolutions over secret-shared values essentially correspond to
the parties running an analogous local operation on their shares, followed by a communication
step (see \cref{sec:crypto}).
Our goal is to take advantage of the GPU to accelerate each party's local computation
on their individual shares. As noted in \cref{sec:architecture-background}, existing
GPU libraries for linear algebra only support computation over 64-bit floating
point values. Thus, to take advantage of GPU support for these operations, we need to embed
the ring operations over $\Z_{2^{64}}$ (or equivalently, 64-bit integer operations)
into 64-bit {\em floating point} operations.

\paragraph{Integer operations using floating-point arithmetic.}
Our approach for embedding 64-integer operations into 64-bit floating point operations
relies on the following observations:
\begin{itemize}[leftmargin=*]
  \item \textbf{Exact computation for small values.}
  First, 64-bit floating point values have 52 bits of precision
  and can exactly represent {\em all} integers in the interval $[-2^{52}, 2^{52}]$. This
  means that for all integers $a, b \in \Z \cap [-2^{26}, 2^{26}]$, we can compute
  the product $ab$ using their floating-point representations and still recover the
  correct value {\em over the integers}.
  
  \item \textbf{Bilinearity.}
  Operations like matrix multiplication and convolutions are {\em bilinear}.
  This means that for any choice of inputs $\Am_1, \Am_2, \Bm_1, \Bm_2$,
  \begin{multline*}
    (\Am_1 + \Am_2) \circ (\Bm_1 + \Bm_2) = \\
     \Am_1 \circ \Bm_1 + \Am_2 \circ \Bm_1 + \Am_2 \circ \Bm_1 + \Am_2 \circ \Bm_2,
  \end{multline*}
  where $\circ$ denotes an arbitrary bilinear operation.
  Suppose now that we rewrite an input as an expansion in a smaller base;
  for example, we might write $\Am = \Am_0 + 2^{16} \Am_1$ and $\Bm = \Bm_0 + 2^{16} \Bm_1$.
  Bilinearity ensures that $\Am \circ \Bm$ can be expressed as a linear
  combination of the pairwise products $\Am_0 \circ \Bm_0$,
  $\Am_0 \circ \Bm_1$, $\Am_1 \circ \Bm_0$, and $\Am_1 \circ \Bm_1$.
  Computing $\Am \circ \Bm$ from the pairwise products
  only requires {\em element-wise}
  additions and scalar multiplications.

  \item \textbf{CUDA kernels for element-wise operations.} To complete the puzzle,
  we note that there are optimized CUDA kernels for performing component-wise addition
  and scalar multiplication on 64-bit {\em integer} values.
\end{itemize}
To evaluate a bilinear operation $\circ$ like matrix multiplication or convolution (which do
{\em not} have integer kernels), $\oursystem$ first decomposes each of the inputs
$\Am, \Bm \in \Z_{2^{64}}^{n \times m}$
into smaller inputs $\Am_1, \ldots, \Am_k, \Bm_1, \ldots, \Bm_k \in \Z_{2^{w}}^{n \times m}$
where $\Am = \sum_{i = 1}^k 2^{(i-1)w} \Am_i$.
Then, it computes the $k^2$ products $\Am_i \circ \Bm_j$ using floating-point arithmetic
on the GPU. As long as the entries of $\Am_i \circ \Bm_j$ do not exceed $2^{52}$ in magnitude,
all of these pairwise products are computed exactly. Finally, each component of the  pairwise product
is re-interpreted as a 64-bit integer. Computing $\Am \circ \Bm$ from the
pairwise products $\Am_i \Bm_j$ amounts to evaluating a linear combination of tensors, which
can be done efficiently using existing CUDA kernels for 64-bit integer operations. Note that
since the final operations are taken modulo $2^{64}$, it suffices to compute only the products
$\Am_i \Bm_j$ where $w(i + j - 2) < 64$.

When performing computations using floating-point kernels,
\oursystem decomposes each input into $k = 4$ blocks, where
the values in each block are represented by a $w = 16$-bit value.
For this choice of parameters, each bilinear operation is expanded into $10$
pairwise products.

\begin{remark}[Smaller Number of Blocks]
  While it may be tempting to decompose 64-bit values into $k = 3$ blocks, where each
  block consists of 22-bit values,
  this compromises correctness of our approach. Namely, correctness of the computation
  is guaranteed only if the entries in 
  each of the intermediate pairwise products $\Am_i \circ \Bm_j$ do not
  exceed the 52-bits of available floating-point precision. If the entries of $\Am_i$ and $\Bm_j$
  are 22 bits, then the entries in a single multiplication between an element in $\Am_i$ and $\Bm_j$
  will already be 44 bits. If we are
  evaluating a convolution (or matrix multiplication) where each output component is a sum of
  $2^8 = 256$ values, this exceeds the available precision and triggers an arithmetic
  overflow. This is problematic for
  larger networks. Using 16-bit blocks, we can handle bilinear operations involving up to
  $2^{20}$ intermediate products, which is sufficient for our applications.
\end{remark}

\begin{remark}[Overhead of Block-wise Decomposition]
  While decomposing each bilinear operation on integer values into $O(k^2)$ floating-point operations
  on same-sized inputs can appear costly, \oursystem takes advantage of GPU parallelism to
  mitigate the {\em computational} overhead. Namely, for convolutions, \oursystem
  uses group convolution (\texttt{cudnnConvolutionForward})
  to compute the
  convolutions in parallel. Similarly, for matrix multiplications, \oursystem uses a
  batch matrix multiplicative kernel (\texttt{cublasSgemm}) to multiply matrices in parallel.
  We observe that for small inputs (e.g., $64 \times 64$ inputs), this approach only incurs
  a modest $2\times$ overhead (compared with evaluating a single convolution of the same size)
  and increases to roughly $9\times$ for larger $224 \times 224$ inputs.

  While the computational overhead of our embedding is partially mitigated through parallelism,
  this approach does increase the memory requirements of our protocol. This does not
  have a significant effect on privacy-preserving
  inference, but
  it does limit the batch size we can handle during privacy-preserving training (recall that
  during training, a single iteration of the optimization algorithm processes a {\em batch} of instances).
  Scaling up to support larger batch sizes during privacy-preserving training would likely necessitate
  distributing the computation across multiple GPUs rather than a single GPU (as is also the case for
  training deep models in the clear).
\end{remark}

\begin{remark}[Comparison with \textsc{Delphi}]
\label{rmk:delphi-float}
  The \textsc{Delphi} system~\cite{MLSZP20} leverage GPUs for evaluating convolutions on secret-shared
  inputs in their private inference system. In their setting, the parameters are chosen so that the outputs
  of the convolution are always within the interval $[2^{-52}, 2^{52}]$, and as such, the existing 
  floating-point kernels for convolution can be used without incurring any floating-point precision issues.
  In particular,
  \textsc{Delphi} uses a 32-bit ring and 15 bits of fixed-point precision. The system works
  in the setting where the model parameters are assumed to be {\em public}: namely, the convolution
  kernels are {\em not} secret-shared. In this way, convolutions are evaluated between a {\em plaintext} value
  and a secret-shared value, which ensures that the
  resulting outputs are bounded. In our setting, both
  the model and the inputs are secret-shared so we cannot directly embed the integer-valued operations
  into 64-bit floating-point computations. In fact, as we discuss in \cref{sec:micro}, to have
  sufficient precision when scaling up to deeper models and larger datasets, it is often necessary to use
  a larger ring (i.e., a 64-bit ring) for the arithmetic secret sharing.
\end{remark}

\paragraph{The \cudalong abstraction.}
\oursystem provides a new abstraction called a \cudalong for embedding
64-bit integer-valued operations into 64-bit floating-point arithmetic. Similar to
\crypten's \mpctensor, the \cudalong abstractly represents a secret-shared
tensor of 64-bit {\em integers} and is backed by a standard
\pytorch tensor of 64-bit integers. In the back end, whenever an elementary operation
needs to be evaluated on the underlying tensor, \oursystem proceeds as follows:
\begin{itemize}[leftmargin=*]
  \item If optimized CUDA kernels exist for evaluating the chosen operation on integer-valued
  tensors (e.g., point-wise addition or point-wise multiplication),
  then the corresponding CUDA kernel is directly invoked.

  \item For bilinear operations where optimized CUDA kernels
  only exist for computations on floating-point inputs
  (e.g., convolutions, matrix multiplications), then \oursystem applies the above
  technique of first decomposing the input into $k = 4$ tensors of 16-bit values,
  computing all necessary $O(k^2)$ pairwise products of the resulting blocks (using the floating point
  kernel), and re-combines the pairwise products to obtain the final output.
\end{itemize}


\section{Threat Model and Cryptographic Design}
\label{sec:crypto}

In this section, we provide a formal specification of our threat model
and a description of the private inference and training functionalities we develop.
We then describe the cryptographic sub-protocols we use to construct our privacy-preserving
training and inference protocols.

We begin by introducing the notation we use in this work. For a
finite set $S$, we write $x \getsr S$ to denote that $x$ is drawn uniform at
random from $S$. We use boldface letters (e.g., $\xv, \yv$) to denote vectors
and use non-boldface letters (e.g., $x_i, y_i$) to denote their components.
We denote our three parties by $P_1, P_2, P_3$. To simplify notation,
whenever we use an index $i \in \{ 1, 2, 3 \}$ to denote a party (or a share), we
write $i - 1$ and $i + 1$ to denote the ``previous'' party and the ``next'' party,
respectively. For example, $P_{3 + 1}$ refers to $P_1$.

We say that a function $f$ is negligible in a parameter $\lambda$ if
$f(\lambda) = o(\lambda^{-c})$ for all $c \in \N$.
We say an algorithm is efficient if it runs in probabilistic
polynomial-time in the length of its input.
We say that two families of
distributions $\calD_1 = \set{\calD_{1, \lambda}}_{\lambda \in \N}$ and
$\calD_2 = \set{\calD_{2, \lambda}}_{\lambda \in \N}$ are computationally
indistinguishable (i.e., $\calD_1 \appc \calD_2$) 
if no efficient adversary can distinguish samples from
$\calD_1$ and $\calD_2$ except with negligible probability.
\subsection{Threat Model}
\label{sec:threat-model}

Similar to several recent 3-party
protocols~\cite{AFLNO16,FLNW17,MR18,WTBKMR21}, we design our system in the
honest-majority model. Moreover, we focus on semi-honest adversaries. Namely,
we assume that each of the three computing parties follow the protocol,
but may {\em individually} try to learn information about other parties'
inputs. Formally, we consider the standard simulation-based notion
of security in the presence of semi-honest adversaries~\cite{Can00,Gol04}:
\begin{definition}[Semi-Honest Security]
  \label{def:semi-honest-sec}
  Let $f \colon (\zo^n)^3 \to (\zo^m)^3$ be a randomized functionality and let
  $\pi$ be a protocol. We say that $\pi$ securely computes $f$
  in the presence of a single semi-honest corruption if
  there exists an efficient simulator $\calS$ such that
  for every corrupted party $i \in \{1, 2, 3\}$ and every
  input $\xv \in (\zo^n)^3$,
  \[  \{ \out^\pi(\xv), \view^\pi_i(\xv) \} \appc \{ f(\xv), \calS(i, x_i, f_i(\xv)) \} \]
  where $\view^\pi_i(\xv)$ is the view of party $i$ in an
  execution of $\pi$ on input $\xv$, $\out^\pi(\xv)$ is the output of all parties
  in an execution of $\pi$ on input $\xv$, and $f_i(\xv)$ denotes the $\ord{i}$
  output of $f(\xv)$.
\end{definition}

\paragraph{Computing on secret-shared values.}
In this work, we consider two main settings: private inference and private training
on secret-shared inputs. We use standard 3-out-of-3 additive secret sharing
as well as 2-out-of-3 replicated secret sharing.
Abstractly, we model both types of secret sharing
as a pair of algorithms $(\Share, \Reconstruct)$ with the following properties:
\begin{itemize}[leftmargin=*]
  \item On input $x \in \zo^n$, the share algorithm
  $\Share(x)$ outputs a tuple of three shares $(x_1, x_2, x_3)$.

  \item The reconstruction algorithm $\Reconstruct(S)$ takes a set of shares
  $T$ and outputs a value $x \in \zo^n$ if successful and $\bot$ if not.
\end{itemize}
Correctness of a threshold secret sharing scheme with threshold $t$
says that for any subset of shares $T \subseteq \Share(x)$
of size at least $t$, $\Reconstruct(T) = x$. Perfect security
says that there exists a probabilistic polynomial-time simulator $\calS$ such that
for every subset $T \subseteq \set{1, 2, 3}$ where $\abs{T} < t$ and every
input $x \in \zo^n$,
\[ \set{(x_1, x_2, x_3) \gets \Share(x) : (x_i)_{i \in T} } \equiv \set{\calS(1^n, T)}. \]
We now formally define our notion of private inference and private training on
secret-shared inputs:
\begin{itemize}[leftmargin=*]
  \item \textbf{Private inference:} Inference is the problem of evaluating
  a trained model $M$ on an input $x$. We denote this operation as $\ms{Eval}(M, x)$.
  In private inference, the ideal functionality $f$
  maps secret shares of an input $x$
  and a model $M$ to a secret share of the output $\ms{Eval}(M, x)$.
  Namely, on input $((M_1, x_1), (M_2, x_2), (M_3, x_3))$, the ideal functionality
  outputs $\Share(\ms{Eval}(M, x))$ where $M \gets \Reconstruct(\set{M_1, M_2, M_3})$ and
  $x \gets \Reconstruct(\set{x_1, x_2, x_3})$.
  In particular, a private inference protocol ensures privacy for the model $M$, the input
  $x$, and the output $\ms{Eval}(M, x)$.

  \item \textbf{Private training:} In private training, the goal is to run a training algorithm
  $\ms{Train}$ on some dataset $D$. In this case, the ideal functionality $f$
  maps secret shares of the dataset $(D_1, D_2, D_3)$ to
  a secret share of the model $\Share(\ms{Train}(D))$ where $D \gets \Reconstruct(D_1, D_2, D_3)$.
  In this case, each party
  individually learn nothing about the input dataset $D$ or the resulting learned model $\ms{Train}(D)$.
\end{itemize}

\subsection{Cryptographic Building Blocks for Private Inference}
\label{sec:crypto-protocols-inf}

We now describe the main MPC building blocks we use for private inference on deep
neural networks. First, we decompose the neural network inference algorithm
into a sequence of elementary operations: linear/pooling/convolution layers
and activation function evaluation (ReLU).
To obtain our protocol $\pi$ for computing the ideal functionality for private
inference, we {\em sequentially} compose the
semi-honest secure protocols for realizing each of the elementary operations. Correctness
and semi-honest
security of the overall protocol then follows by correctness and 
security of the underlying sub-protocols
together with the sequential composition theorem~\cite{Can00}.

\paragraph{``GPU-friendly'' cryptography.} As alluded to in \cref{sec:our-contributions,sec:sys-design},
we seek cryptographic protocols that are particularly amenable to GPU acceleration.
For example, protocols that involve conditionals (such as garbled circuits~\cite{Yao86})
or require extensive finite field arithmetic are more challenging to support efficiently
on the GPU. For this reason, we focus primarily on secret-sharing based protocols
and work over a ring with a power-of-two modulus. In the following description, we elect
to use cryptographic protocols where the underlying implementations vectorize and whose
evaluation can be expressed primarily in terms of point-wise or component-wise operation
on blocks of data.

\paragraph{Secret sharing.} We work over the ring
$\ring$ where $n =\ 2^k$ is a power of 2. In our specific implementation,
$k = 64$. To secret share a value
$x \in \ring$, sample shares $x_1, x_2, x_3 \getsr \ring$ such that
$x_1 + x_2 + x_3 = x$. Following Araki~\etal~\cite{AFLNO16}, our default
sharing is a 2-out-of-3 ``replicated secret sharing''~\cite{ISN89} where
each party holds a pair of shares: $P_1$ has $(x_1, x_2)$, $P_2$ has
$(x_2, x_3)$, and $P_3$ has $(x_3, x_1)$. We denote
this by $\share{x} = (x_1, x_2, x_3)$. In some cases,
we will also consider a 3-out-of-3 additive secret sharing scheme where
party $P_i$ holds $x_i$ (but {\em none} of the other shares).

\paragraph{Fixed point representation.} Machine learning algorithms natively operate on
real (i.e., floating-point) values while the most efficient cryptographic 
protocols are restricted to computations over discrete
domains such as rings and finite fields. Following previous work,
we use a fixed-point encoding of all values occurring in the computation,
and then embed the integer-valued
fixed-point operations in the ring $\ring$. Specifically,
if we consider a fixed-point encoding with $t$ bits of precision, a real value
$x \in \R$ is represented by the integer $\lfloor x \cdot 2^{t} \rceil$ (i.e., the nearest
integer to $x \cdot 2^{t}$). The ring modulus $n$ is chosen to ensure
no overflow of the integer-valued fixed-point operations. \oursystem
sets $n = 64$; we discuss this choice in detail in \cref{sec:acc}.

\paragraph{Protocol initialization.} In the following description, we assume
that the parties have many independent secret shares of $0$. This will
be used for ``re-randomization'' during the protocol execution. We implement this using
the approach of Araki~\etal~\cite{AFLNO16}. Specifically, let 
$F$ be a pseudorandom function (PRF).
At the beginning of the protocol,
each party $P_i$ samples a PRF key $k_i$ and sends $k_i$ to
party $P_{i + 1}$. The $\ord{j}$ secret share of $0$ is the triple
$(z_1, z_2, z_3)$ where $z_i = F(k_i, j) - F(k_{i - 1}, j)$.

\paragraph{Linear operations.} Linear operations on secret-shared data only
require local computation. Namely, if $\alpha, \beta, \gamma \in \ring$ are public
constants and $\share{x}, \share{y}$ are secret-shared values, then
$\share{\alpha x + \beta y + \gamma} =
  (\alpha x_1 + \beta y_1 + \gamma, \alpha x_2 + \beta y_2, \alpha x_3 + \beta y_3)$.
Each of the parties can compute their respective shares of $\share{\alpha x + \beta y + \gamma}$
from their shares of $\share{x}$ and $\share{y}$
and the public coefficients $\alpha, \beta, \gamma$.

\paragraph{Multiplication.} To multiply two secret-shared values
$\share{x} = (x_1, x_2, x_3), \share{y} = (y_1, y_2, y_3)$,
each party $P_i$ locally computes $z_i = x_i y_i + x_{i + 1} y_i + x_i y_{i + 1} $.
By construction, $z_1 + z_2 + z_3 = xy \in \ring$. This yields a 3-out-of-3 additive
sharing of $z$. To obtain replicated shares of $z$, party $P_i$ sends
$P_{i + 1}$ a blinded share $z_i + \alpha_i$, where $(\alpha_1, \alpha_2, \alpha_3)$
is a fresh secret-sharing of $0$ (derived from the PRF as described above).

Since $x, y$ are fixed-point encodings, the parties additionally need to {\em rescale}
$z$ after computing the product (i.e., divide it by the scaling factor $2^t$). In this work,
we use the truncation protocol $\Pi_{\mathsf{trunc1}}$ from \aby~\cite{MR18} to implement this procedure.
We note that Mohassel and Rindal propose two versions of the share truncation protocol: a two-round protocol
$\Pi_{\mathsf{trunc1}}$ that only relies on elementary arithmetic operations and a one-round protocol
$\Pi_{\mathsf{trunc2}}$ that relies on precomputed ``truncation tuples''. While generating the truncation
tuples can be done in a separate offline phase, doing so requires implementing a Boolean bit extraction
circuit over secret-shared values. In contrast, $\Pi_{\mathsf{trunc1}}$ relies exclusively on 
arithmetic operations, and naturally extends to our tensor-based computing model. For this reason,
we use the two-round truncation protocol $\Pi_{\mathsf{trunc1}}$ in our implementation.
This has the added advantage that we avoid
a separate (and potentially expensive) preprocessing step.
Both of these share-truncation protocols are not exact and may introduce $1$
bit of error in the {\em least} significant bit of the secret-shared value (i.e., with $t$ bits of
fixed-point precision, the error introduced is bounded by $2^{-t}$). We provide an empirical
assessment of the error (and resulting model accuracy) in \cref{sec:acc}.

\paragraph{Convolutions and matrix multiplication.} The above protocols for computing linear functions as well
as products of secret-shared values directly vectorize to yield
protocols for computing linear functions on tensors as well as bilinear operations like matrix
multiplication and convolution. Linear functions on secret-shared tensors only require
local computation. Bilinear operations on secret-shared tensors like matrix multiplications and 
convolutions are implemented by computing three
separate products (as described in the multiplication protocol above).
These computations over secret-shared tensors directly map to analogous computations
on local shares, so we can take advantage of existing highly-optimized CUDA
kernels for evaluating these operations via the technique from \cref{sec:sys-design}.

As in several previous systems (e.g.,~\cite{MZ17, MR18, KRCGRS20}),
when we compute products of secret-shared tensors, we only
apply the truncation protocol to the {\em result} of the product
and {\em not} after each
individual multiplication.
This has a significant impact on the performance of the protocol for two
reasons: (1) we can use existing CUDA kernels optimized for matrix products and convolutions without
needing to modify how the elementary multiplications are performed; and (2)
the total communication in the protocol is proportional to the size of the {\em output}
rather than the number of {\em intermediate} element-wise multiplications.

\paragraph{Most significant bit.} Several of our protocols rely on a protocol
for computing the most significant bit $\share{\mathrm{msb}(x)}$
of a secret-shared value $\share{x}$. In our fixed-point representation,
this corresponds to computing the sign of $x$. For
this, we adopt the general approach from \aby. Namely,
given an arithmetic secret sharing $\share{x} = (x_1, x_2, x_3)$ of $x$, the parties
re-interpret it as three {\em binary} shares of values $\bshare{x_1} = (x_1, 0, 0)$,
$\bshare{x_2} = (0, x_2, 0)$, and $\bshare{x_3} = (0, 0, x_3)$. The parties now evaluate
an addition circuit on the binary shares $\bshare{x_1}, \bshare{x_2}, \bshare{x_3}$
to compute binary shares of the sum $\bshare{x}$, which in particular, yields a binary
share of $\bshare{\mathrm{msb}(x)}$. Finally, to recover arithmetic shares of $\share{\relu(x)}$
from $\share{x}$ and $\bshare{\mathrm{msb}(x)}$, we use the bit injection
protocol from $\aby$~\cite[\S5.4]{MR18}, which only requires simple arithmetic operations.

The majority of this computation is the evaluation of the addition
circuit over binary shares on the GPU. 
Evaluating a Boolean addition circuit on secret-shared binary values decomposes into a
sequence of bitwise \textsc{and} and \textsc{xor} operations
(along with communication for the \textsc{and} gates), which can be computed using
efficient GPU kernels.
We provide microbenchmarks in \cref{sec:micro}.

\paragraph{ReLU activation function.} The standard activation function we consider
in our networks is the rectified linear unit (ReLU)~\cite{NH10,KSH12}:
$\relu(x) \deq \max(x, 0)$. To compute the ReLU function, it suffices to construct
a protocol for testing whether the fixed-point value $x$ is positive or not. This corresponds
to computing the most significant bit $\mathrm{msb}(x)$ of $x$, which we evaluate
using the protocol described above.

\iftoggle{fullversion}{

\iftoggle{fullversion}{
\subsection{Additional Building Blocks for Private Training}
}{
  \section{Building Blocks for Private Training}
}
\label{sec:crypto-protocols-train}
To support private training, we need to augment our existing toolkit with
several additional protocols. Here, we consider a standard backpropagation setting
with a softmax/cross-entropy loss function optimized using (minibatch)
stochastic gradient descent (SGD)~\cite{GBC16}.
As with private inference, we decompose the backpropagation algorithm into
a sequence of elementary operations
and build our private training protocol by sequentially
composing protocols for the elementary operations.

In this work, we consider classification tasks with $d$ target classes.
Each iteration of SGD takes an input $\xv \in \R^m$ and a one-hot encoding of the
target vector $\yv \in \zo^d$ (i.e., $y_i = 1$ if $\xv$ belongs to class $i$ and $y_i = 0$ otherwise)
and computes the cross-entropy loss:\footnote{Technically, in minibatch SGD, each
iteration takes a batch of $N$ inputs and the loss function is the average of
the loss function for all $N$ inputs in the batch. For ease of exposition, we describe
the setup for a single input, but everything generalizes naturally to the minibatch setting.}
\begin{equation}
  \label{eq:cross-entropy}
  \ell_{\ms{CE}}(\xv ; \yv) \deq - \sum_{i \in [d]} y_i \log \tilde z_i,
\end{equation}
where $\tilde \zv \gets \softmax(\zv)$, $\zv \gets \ms{Eval}(M, \xv)$,
and $M$ is the current model.
For a vector $\xv \in \R^d$, the softmax function
\begin{equation}
  \label{eq:sofmax}
  \softmax_i(\xv) \deq e^{x_i} / \sum_{i \in [d]} e^{x_i}.
\end{equation}
The gradient of $\ell_{\ms{CE}}$
for the output layer $\zv$ is then
\[ \nabla_{\zv} \ell_{\ms{CE}} = \softmax(\zv) - \yv. \]
We can use the private inference protocol from
\cref{sec:crypto-protocols-inf} to
compute $\share{\zv}$ from $\share{\xv}$ and $\share{M}$.
To compute $\share{\nabla_{\zv} \ell_{\ms{CE}}}$, we
need a protocol to compute softmax on secret-shared values.

For the ReLU layers, the gradient computation reduces to evaluating
the derivative of the ReLU function. The gradients
for the linear/convolution layers are themselves linear
functions of the gradients from the preceding layer, and thus, can be handled
using the protocols from \cref{sec:crypto-protocols-inf}. In the following,
we describe our protocols for evaluating the softmax and the derivative of the
ReLU function on secret-shared values.
Note that backpropagation does {\em not} require computing the value of the loss function (\cref{eq:cross-entropy}),
so we do {\em not} need a protocol for computing logarithms on secret-shared values.

\paragraph{Softmax.} To avoid numeric
imprecision from evaluating the exponential function in the softmax function (\cref{eq:sofmax}) on
very large or very small inputs, a standard technique is to evaluate the softmax
on the ``normalized'' vector $(\xv - \max_i x_i)$~\cite{GBC16}. A simple calculation
shows that $\softmax(\xv - \max_i x_i) = \softmax(\xv)$. This has the advantage
that all inputs to the exponential function in \cref{eq:sofmax} are at most $0$, and the
denominator is contained in the interval $[1, d]$.
In the following, we describe protocols for evaluating the exponential function,
division, and computing the maximum over a vector of secret-shared values. Together,
this yields a protocol for computing a softmax on secret-shared values.

\paragraph{Exponentiation.} We approximate the exponential function $e^x$ needed
to compute softmax with its limit characterization $f_m$:
\begin{equation}
  \label{eq:exp-approx}
  f_m(x) \deq \left(1 + \frac{x}{m} \right)^m.
\end{equation}
Using a Taylor expansion for the function $\ln(1 + x)$ and assuming that
$\abs{x} < m$,
\[ \frac{f_m(x)}{e^x} = \frac{e^{m \ln(1 + x/m)}}{e^{x}} = e^{-O(x^2/m)}. \]
Thus, the degree-$m$ approximation $f_m$ provides a good approximation
$e^x$ on an interval of size $O(\sqrt{m})$ centered at $0$. A common alternative approximation
is to use Taylor series to approximate the exponential function.
The advantage of using a Taylor series approximation of degree $m$ is that it provides
a good estimate in an interval of size $O(m)$ (as opposed to $O(\sqrt{m})$ using
the approximation $f_m$). However, using a Taylor series approximation has several
drawbacks:
\begin{itemize}[leftmargin=*]
\item Evaluating a degree-$m$ Taylor approximation requires $m$ multiplications
over $O(\log m)$ rounds. Computing $f_m$, in comparison, 
only requires $\log m$ multiplications.
For a fixed degree $m$, 
the cost of computing the $f_m$ approximation is {\em exponentially}
smaller than computing the degree-$m$ Taylor series approximation.

\item The size of the smallest coefficient in the Taylor series of degree $m$ is
$1/m!$. In a fixed-point encoding scheme with $t$ bits of precision, values less than
$2^{-t - 1}$ round to $0$. This gives an upper bound on the degree of the Taylor expansion
we can feasibly support. Alternatively, we could compute the terms $x^m/m!$ in the Taylor expansion as
$\prod_{i \in [m]} \frac{x}{i}$, but this now requires $O(m)$ {\em rounds} of multiplications
to compute.

\item In our setting, the inputs to the exponential function are drawn from the interval $(-\infty, 0]$.
The approximation $f_m(x)$ has the appealing property that as $x \to -\infty$, $f(x) \to 0$, which
matches the behavior of $e^x$. In contrast, the Taylor approximation {\em diverges} as $x \to -\infty$.
This can introduce significant errors in the computation (unless we use a Taylor approximation of
sufficiently high degree). For the models and inputs we consider in \cref{sec:exp-setup},
most inputs to the exponential function lie in the interval $[-45, 0]$. Ensuring that the Taylor
approximation does not diverge for all inputs in this interval would require a high-degree approximation.

\end{itemize}
Thus, compared to a Taylor approximation,
the limit-based approximation $f_m$ is more efficient to evaluate (in terms of the number
of multiplications) and more robust for handling large negative inputs that may arise in
the computation.

\paragraph{Division.} Computing the softmax function requires
computing a quotient $\share{x / y}$ on secret-shared values $\share{x}$ and $\share{y}$
and where $1 \le y \le Y$, for some bound $Y$.
It suffices to compute the reciprocal $\share{1 / y}$ and compute the quotient
using share multiplication. Similar to previous works~\cite{WTBKMR21},
we use the iterative Newton-Raphson
algorithm to approximate the value of $1/y$. Very briefly, the
Newton-Raphson algorithm for approximating $1 / y$
starts with an initial ``guess'' $z_0$ and iteratively
computes $z_i \gets 2 z_{i-1} - yz_{i-1}^2$. In this work, we use a
{\em fixed} initialization $z_0 = 1 / Y$. This provides a highly-accurate
estimate for $1/y$ for all $y \in [1, Y]$ using $O(\log Y)$ iterations
of Newton's algorithm. To see this, let
\[ \text{error}_i = \abs{1/y - z_i} = \frac{1}{y} \abs{1 - z_i y} \le \varepsilon_i, \]
where $\varepsilon_i = \abs{1 - z_i y}$. Substituting in the Newton-Raphson updates,
$\varepsilon_i = \varepsilon_{i - 1}^2$, so the {\em maximum}
error after $i$ iterations is $(1 - 1/Y)^{2^i} \le e^{-2^i / Y}$.

We note that using a more accurate initialization for Newton-Raphson will allow convergence
in fewer iterations. However, methods
for computing a more accurate estimate~\cite{WTBKMR21} for the initialization
typically rely on binary-valued operations (e.g., comparisons) and are {\em more}
costly than using a fixed initialization and increasing the number of iterations.
Note that a fixed initialization is possible in our setting because we are guaranteed
that the values $y$ lies in a fixed interval (due to the normalization in the softmax computation).

\paragraph{Maximum.} The last ingredient we require for computing the softmax
function is computing the maximum value $\share{\max_i x_i}$ from a secret-shared
vector $\share{\xv}$ where $\xv \in \R^m$.
We implement this using $m$ invocations of a comparison protocol. To reduce
the round complexity to $\log m$, we use a tree of comparisons where
pairs of elements are compared each round, and the
larger value in each pair advances to the next round.
Comparing two secret-shared {\em fixed-point}
values $\share{x}, \share{y}$ is equivalent
to computing the most significant bit of their difference
(i.e., $\share{\mathrm{msb}(x - y)}$). We implement this
using the protocol from \cref{sec:crypto-protocols-inf}.

\paragraph{Derivative of ReLU.} During backpropagation, we also need
to compute the derivative of the ReLU function $\relu'(x)$, which is $0$
if $x < 0$ and $1$ if $x > 0$. This again corresponds to computing the
most significant bit of the fixed-point encoding of $x$, which we implement
using the protocol from \cref{sec:crypto-protocols-inf}.

}{
  \paragraph{Privacy-preserving training.} In \cref{sec:crypto-protocols-train}, we
  describe additional protocols we use for privacy-preserving training.
}


\section{System Implementation and Evaluation}
\label{sec:exp}

We build \oursystem
on top of \crypten, which itself builds on \pytorch.
First, we introduce the \cudalong data type that represents a \pytorch
tensor for 64-bit integer values (see \cref{sec:sys-design}). Our design enables
us to take advantage of optimized CUDA kernels for evaluating bilinear operations
such as convolutions and matrix multiplications on secret-shared tensors. This suffices
for evaluating arithmetic circuits on secret-shared tensors. Using
these elementary building blocks, we then implement protocols for each of the
operations described in \cref{sec:crypto} (i.e.,
the truncation protocol for fixed-point multiplication, ReLU computation,
and the softmax function). Through composing these individual protocols together,
we obtain an end-to-end system for private inference and private
training.

\paragraph{Point-to-point communication.} We leverage \pytorch's \texttt{torch.distributed} package for point-to-point communication between parties. The default communication mode in \pytorch is a ``broadcast'' mode where
every message sent by a party is sent to {\em all} peers. To emulate point-to-point channels
(as required by our protocol), we initialize a separate communication back end between each pair
of parties. In this case, a ``broadcast'' channel between each pair of parties functions
as a point-to-point channel between the parties. 

\paragraph{Pseudorandom generators on the GPU.} We use AES as the PRF in our protocol
(used for share re-randomization in the truncation protocol). We
use the \texttt{torchcsprng} \pytorch C++/CUDA extension~\cite{PyTorchAES}
(based on the Salmon~\etal protocol~\cite{SMDS11}) which enables
AES evaluation on the GPU. 

\subsection{Experimental Setup for System Evaluation}
\label{sec:exp-setup}

We now describe our experimental setup for evaluating \oursystem
as well as the specific parameters we use to instantiate our
cryptographic protocols from \cref{sec:crypto}.
\iftoggle{fullversion}{

\iftoggle{fullversion}{}{
  \section{Datasets and Models}
  \label{sec:datasets}
  In this section, we describe the datasets and machine learning models we use in our evaluation.
}

\paragraph{Deep learning datasets.}
We evaluate \oursystem on the following standard datasets for object recognition:
\begin{itemize}[leftmargin=*]
  \item \textbf{MNIST}~\cite{MNIST}. MNIST is a dataset for handwritten digit recognition. The training set has 60,000 images and the test set has 10,000 images. Each digit is a grayscale (i.e., single-channel) $28\times 28$ image.
  Due to its relatively small size, it is widely used as a benchmark in many privacy-preserving ML systems~\cite{MZ17, MR18, KRCGRS20, WTBKMR21}.

  \item \textbf{CIFAR-10}~\cite{CIFAR}. CIFAR-10 is a dataset with 60,000 $32 \times 32$ RGB images split evenly across 10 classes.
  
  \item \textbf{Tiny ImageNet}~\cite{TinyImageNet}. Tiny ImageNet is a modified subset of the ImageNet dataset. It contains 100,000 $64 \times 64$ RGB training images and 10,000 testing images split across 200 classes.
  Compared to CIFAR-10, Tiny ImageNet is much more challenging: each image is $4\times$ larger
  and there are $20\times$ more classes.
  
  \item \textbf{ImageNet}~\cite{ImageNet}. ImageNet is a large-scale visual recognition dataset with more than 
  1,000,000 training images. It is the standard benchmark for evaluating the classification performance of computer vision models. ImageNet has 1000 classes, and each example is a center-cropped $224 \times 224$ RGB image. The only
  prior system for privacy-preserving machine learning
  that demonstrates performance at the scale of ImageNet is \textsc{CrypTFlow}~\cite{KRCGRS20}.
\end{itemize}

\paragraph{Deep learning models.}
For our experimental evaluation, we measure the cost of our private training and private inference protocols
on several representative CNN architectures developed for object recognition.
Each of these networks can be represented as a composition of a collection of
standard layers: convolution, pooling, activation, batch normalization, softmax,
and fully-connected layers.

\begin{itemize}[leftmargin=*]
  \item \textbf{LeNet}~\cite{LBBH98}. LeNet was proposed by
  LeCun~\etal for handwritten digit recognition. It is a
  shallow network with 2 convolutional layers, 2 average pooling
  layers, and 2 fully connected layers. The network uses the hyperbolic tangent
  ($\tanh$) as its activation function.

  \item \textbf{AlexNet}~\cite{KSH12}. AlexNet was the winner of 2012 ImageNet Large Scale Visual Recognition Challenge (ILSVRC-2012) competition. It has 5 convolutional layers, 3 max pooling layers, and 2 fully connected layers for
  a total of 61 million parameters. 
  AlexNet uses ReLU as its activation function.

  \item \textbf{VGG-16}~\cite{SZ14}. VGG-16 is the runner-up of the ILSVRC-2014 competition. It uses 16 layers consisting of convolution, ReLU, max pooling, and fully-connected layers.
  VGG-16 has a total of 138 million parameters.

  \item \textbf{ResNet}~\cite{HZRS16}. ResNet is the winner of ILSVRC-2015 competition. It introduces skip-connections that addresses the vanishing gradient problem when training deep neural network models. ResNet consists of convolution, max pooling, average pooling, batch normalization, and fully connected layers. Since their inception,
  the ResNet family of models have enjoyed wide adoption in the computer vision community.
  We evaluate the performance of ResNet-50, ResNet-101, and ResNet-152 on ImageNet. These networks respectively
  have 23, 44, and 60 million parameters and 50, 101, and 152 layers.
\end{itemize}

\paragraph{Architecture adjustments.} We use the standard architecture of each of these networks,
except with the following modifications:
\begin{itemize}[leftmargin=*]
  \item \textbf{AlexNet and VGG-16 on small datasets.} 
  Since AlexNet and VGG-16 were designed for ImageNet, they are not directly compatible with smaller inputs (i.e., those from CIFAR-10 or Tiny ImageNet).
  Thus, when using AlexNet or VGG-16 with smaller inputs,
  we have to modify the network architecture.
  For AlexNet, we drop the final max pooling layer for CIFAR-10, and adjust
  the number of neurons in the fully-connected classification layers to
  $256$-$256$-$10$ and $1024$-$1024$-$200$ for CIFAR-10 and Tiny ImageNet, respectively.
  For VGG-16, we adjust the number of neurons in the fully-connected classification layers
  to $256$-$256$-$10$ and $512$-$512$-$200$ for CIFAR-10 and Tiny ImageNet, respectively.\footnote{Previous
  systems like \textsc{Falcon}~\cite{WTBKMR21} made similar adjustments when evaluating AlexNet and
  VGG-16 on smaller datasets.}
  When evaluating AlexNet on ImageNet, we use the original architecture~\cite{KSH12}.
  In the case of VGG-16, we add a 2x2 average pooling layer to reduce
  the input dimension of the first fully connected layer from $18432$ to $4608$;
  this is due to memory limitations on the GPU. When we compare our system to
  the \textsc{Falcon} system on these models and datasets, we
  make the same adaptations. We provide the full specification of the AlexNet and VGG-16
  model architectures we use in \cref{sec:network-arch}.

  \item \textbf{Activation functions.} All networks we consider except
  LeNet use the ReLU function as the activation function. In contrast, LeNet uses the hyperbolic
    tangent function $\tanh$ as the underlying activation function. Since \oursystem does not
    support evaluating the $\tanh$ function and modern networks primarily use ReLU as their activation
    function, we replace $\tanh$ with ReLU in our experiments with LeNet.

  \item \textbf{Average pooling.} Pooling is a standard way to down-sample the outputs of the
  convolutional layers in a CNN. Specifically, a pooling layer accumulates the
  output of the convolutional layers by replacing each (small) window of the
  feature map (from the convolutional layer) with the average of the values
  (i.e., average pooling) or the max of the values (i.e., max pooling). Earlier networks
  such as AlexNet and VGG-16 used max pooling throughout, while more recent deep networks
  such as the ResNets primarily use average pooling (with a {\em single} max pooling layer
  at the beginning). While the choice of pooling function does not make
  a significant difference in the computational costs of plaintext training, this is not the
  case in private training. The difference is due to the fact that
  average pooling is a {\em linear} operation
  while max pooling is a highly {\em non-linear} operation. To reduce the
  computational overhead of our system, we replace
  all the max pooling layers in the above networks with average pooling. This
  reduces the complexity at the cryptographic level and allows
  us to take better advantage of GPU parallelism.

  \medskip

  We show in \cref{sec:benchmarks-main} that in existing systems, the pooling
  layer is {\em not} the bottleneck, and the performance improvements of our
  protocol relative to past works is {\em not} due to our substitution of
  average pooling in place of max pooling. We additionally show in \cref{sec:acc}
  that using average pooling in place of max pooling does
  not significantly affect the accuracy of the models we consider.
\end{itemize}
}{We describe the specific datasets and models
we consider in \cref{sec:datasets}.}

\paragraph{Protocol instantiation.} We instantiate our protocols from \cref{sec:crypto}
using the following parameter settings:
\begin{itemize}[leftmargin=*]
  \item \textbf{Fixed-point precision}. We consider secret-sharing schemes
  over the 64-bit ring $\Z_{2^{64}}$, and encode inputs using
  a fixed-point representation with $t = 20$ bits of fractional precision (i.e.,
  an input $x \in \R$ is encoded as $\lfloor x \cdot 2^{20} \rceil$.
  In \cref{sec:acc}, we analyze the effect
  the number of bits of precision has on the accuracy of our protocols.

  \item \textbf{Exponentiation.} We use the function $f_m$ from \cref{eq:exp-approx}
  to approximate the exponential function. In this work, we take $m = 2^9 = 512$, so
  evaluating $f_m$ requires $\log m = 9$ rounds of multiplication.
  With $t = 20$ bits of fixed-point precision, we measure the maximum error 
  of our approximation on all inputs $x \le 0$ to be at most $6 \cdot 10^{-4}$.

  \item \textbf{Division.} As described in \cref{sec:crypto-protocols-train},
  we require a private division protocol to compute $\share{1/y}$ where $y \in [1, Y]$,
  and $Y$ is the number of classes in the classification problem. For all of the datasets
  we consider for private training, $Y \le 200$. In our implementation,
  we use 13 iterations of Newton-Raphson (with $1/Y$ as the initialization).
  With $t = 20$ bits of fixed-point precision, we measure the maximum
  absolute difference between the approximate
  value and the true value for inputs in the interval $[1, Y]$
  to be $\approx 10^{-4}$ (and $\approx 10^{-9}$ using floating-point evaluation).
\end{itemize}

\subsection{Benchmarks for Private Training and Inference}
\label{sec:benchmarks-main}
We run our experiments on three Amazon EC2
instances optimized for {\em GPU computation} (\texttt{p3.2xlarge}). Each
instance has a single NVIDIA Tesla V100 GPU with 16 GB of GPU memory. All of
the instances run Ubuntu 18.4 and have 8 Intel Xeon E5-2686 v4 (2.3 GHz) CPUs and 61 GB of RAM.
We consider a local area network (LAN) environment and
place all three servers in the \texttt{us-east-1} (Northern Virginia) region.
In this case, we measure the network bandwidth to be 1.25GB/s with an average latency of 0.2ms. 
For each model/dataset pair we consider in our evaluation, we measure the end-to-end
protocol execution time and the total amount of communication.

\paragraph{Comparisons with prior work.}
We compare the
performance of \oursystem
against \textsc{Falcon}~\cite{WTBKMR21} and \textsc{CrypTFlow}~\cite{KRCGRS20}.
To our knowledge,
these are the only privacy-preserving machine-learning 
frameworks that have demonstrated the ability to handle neural networks at the scale of AlexNet
on large datasets. Since our primary focus is on the scalability of our approach and not on the performance
on shallow networks (where GPUs are unlikely to shine compared to optimized CPU protocols),
we focus our comparisons with \textsc{Falcon} and \textsc{CrypTFlow}.
\begin{itemize}[leftmargin=*]
  \item For \textsc{CrypTFlow} (which supports private inference for ResNet),
  we use the performance numbers reported in their paper (which also operate
  in a LAN environment). 

  \item For \textsc{Falcon} (which supports private inference and private training for LeNet, AlexNet, and VGG-16),
  we collect benchmarks using their provided reference
  implementation~\cite{FalconCode}.
  We run the \textsc{Falcon} system on three {\em compute-optimized}
  AWS instances (\texttt{c4.8xlarge}) in the Northern Virginia
  region.\footnote{Note that we use {\em different} instances for our comparison because
  \oursystem is {\em GPU-based} while \textsc{Falcon} is {\em CPU-based}.} Each instance runs Ubuntu 18.4 and has
  36 Xeon E5-2666 v3 (2.9 GHz) CPUs and 60~GB of RAM.
  We measure the network
  bandwidth between machines to be 1.16GB/s with an average latency of 0.2ms.
\end{itemize}
For the main benchmarks, we also measure the computational cost using \pytorch
on {\em plaintext} data (with GPU acceleration).

\paragraph{Private inference.} \cref{tab:inference-comp} summarizes the performance
of \oursystem's private inference protocol on the models and datasets described in
\cref{sec:exp-setup}. For
shallow networks and small datasets (e.g., LeNet on MNIST or AlexNet on CIFAR),
\textsc{Falcon} outperforms \oursystem. However, as we scale to progressively larger datasets
and deeper models (e.g., VGG-16 on Tiny ImageNet), then \oursystem is
faster ($3.7\times$ on \mbox{VGG-16}). The performance on small
datasets is not unexpected; after all, if the computation is sufficiently simple, then the extra
parallelism provided by the GPU is unlikely to benefit. Moreover, the use of more efficient
cryptographic building blocks (which may not be ``GPU-friendly'') can allow a CPU-based approach
to enjoy superior performance.

The setting where we would expect the GPU-based approach to perform well is in the setting of large datasets
and deeper models. For instance, at the scale of ImageNet, \oursystem is able 
to perform private inference over the ResNet-152 network
(containing over 150 layers and over 60 million parameters)
in just over 25 seconds. This is about $2.2\times$ faster than \textsc{CrypTFlow}, which
to our knowledge, is the only protocol for private inference that has demonstrated support for
the ResNet family of networks on the ImageNet dataset. For the ResNet family
of networks, the running time of \oursystem scales linearly with the depth of the network.

Compared to plaintext inference on the GPU, there still remains a significant
$1000\times$ gap in performance. This underscores the importance of designing
more GPU-friendly cryptographic primitives to bridge this gap in performance.

\begin{table*}[t]
  \begin{minipage}[t]{\linewidth}
  \renewcommand{\thempfootnote}{\fnsymbol{mpfootnote}}
  \renewcommand{\footnoterule}{}
  \scriptsize
  \centering
  \begin{tabular}{lcc c cc c cc c cc c cc}
    \toprule
    & \multicolumn{2}{c}{{\bf LeNet (MNIST)}} &&
      \multicolumn{2}{c}{\bf AlexNet (CIFAR)} &&
      \multicolumn{2}{c}{\bf VGG-16 (CIFAR)}  &&
      \multicolumn{2}{c}{\bf AlexNet (TI)}    &&
      \multicolumn{2}{c}{\bf VGG-16 (TI)} \\
      \cmidrule{2-3} \cmidrule{5-6} \cmidrule{8-9} \cmidrule{11-12} \cmidrule{14-15}
    & Time & Comm. (MB) && Time & Comm. (MB) && Time & Comm. (MB) && Time & Comm. (MB) && Time & Comm. (MB) \\ \midrule
    {\bf \textsc{Falcon}} & 0.038 & 2.29  && 0.11 & 4.02  && 1.44 & 40.45 && 0.34 & 16.23 && 8.61 & 161.71 \\ 
       \rowcolor{yellow!40!white}
    {\bf \oursystem}      & 0.38  & 3.00  && 0.91 & 2.43  && 2.14 & 56.2  && {0.95} & {13.97} && {2.30} & 224.5 \vspace{0.2em} \\ \midrule 
    {\bf Plaintext} & 0.0007 & --- && 0.0012 & --- && 0.0024 & --- && 0.0012 & --- && 0.0024 & ---\\ \midrule[1.8\cmidrulewidth]

    & \multicolumn{2}{c}{{\bf AlexNet (ImageNet)}}  &&
      \multicolumn{2}{c}{\bf VGG (ImageNet)}        &&
      \multicolumn{2}{c}{\bf ResNet-50 (ImageNet)}  &&
      \multicolumn{2}{c}{\bf ResNet-101 (ImageNet)} && 
      \multicolumn{2}{c}{\bf ResNet-152 (ImageNet)} \\
      \cmidrule{2-3} \cmidrule{5-6} \cmidrule{8-9} \cmidrule{11-12} \cmidrule{14-15}
    & Time & Comm. (GB) && Time & Comm. (GB) && Time & Comm. (GB) && Time & Comm. (GB) && Time & Comm. (GB) \\ \midrule
    {\bf \textsc{CrypTFlow}} & ---    & ---    && ---   & ---    && 25.9        & 6.9         && 40\footnote{Value estimated from \cite[Fig.~10]{KRCGRS20}}
      & 10.5\mpfootnotemark[1]    && 60\mpfootnotemark[1]         & 14.5\mpfootnotemark[1]   \\ 
       \rowcolor{yellow!40!white}
    {\bf \oursystem} & 1.52  & 0.24  && 9.44 & 2.75  && {9.31} & {3.08} && 17.62 & 4.64 && 25.77 & 6.56  \\ \midrule
    {\bf Plaintext} & 0.0013 & --- && 0.0024 & --- && 0.011 & --- && 0.021 & --- && 0.031 & --- \\ \bottomrule
  \end{tabular}
  \end{minipage}
  \caption{Running time (in seconds) and total communication
           of private inference for different models, datasets, and systems in a LAN setting.
           The ``TI'' dataset refers to the Tiny ImageNet dataset~\cite{TinyImageNet}.
           The plaintext measurements correspond to the cost of inference on plaintext data on the GPU (using \pytorch).
           Performance numbers for \textsc{CrypTFlow} are taken from~\cite{KRCGRS20}. Performance numbers for \textsc{Falcon} are
           obtained by running its reference implementation~\cite{FalconCode} on three compute-optimized AWS instances in a LAN environment (see \cref{sec:benchmarks-main}). As discussed in \cref{sec:exp-setup}, when testing the performance of \textsc{CryptGPU}, we replace max pooling with average pooling in 
           all of the networks. }
  \label{tab:inference-comp}
\end{table*}

\paragraph{Batch private inference.} We can also leverage GPU parallelism to process
a {\em batch} of images. This allows us to amortize the cost of private inference.
\cref{tab:amortized-cifar}
shows the time and communication needed for private inference over a batch of
64 images on the CIFAR-10 dataset. Here, the amortized cost of private
inference on a single image using AlexNet drops from 0.91s to 0.017s (a $53\times$ reduction).
With VGG-16, batch processing reduces the per-image cost from 2.14s to 0.18s (a $12\times$
reduction).

\cref{tab:amortized-imagenet} shows the time and communication needed
for private inference on ImageNet using the ResNet networks with a batch of 8 images. Here,
we see a $1.9\times$ reduction in the amortized per-image private inference cost for
each of \mbox{ResNet-50}, ResNet-101, and ResNet-152. The cost reduction compared to
those on the CIFAR-10 dataset (\cref{tab:amortized-cifar}) is smaller. This is likely
due to the smaller batch sizes in play here (8 vs. 64). Supporting larger batch sizes is possible
by either using multiple GPUs or using GPUs with more available memory.
Nonetheless, irrespective of the model/input size, we observe that batch private inference
allows us to amortize the cost of private inference
protocol. Communication in all cases scales linearly with the batch size.

\begin{table}
  \centering
  \begin{tabular}{lcc c cc}
    \toprule 
    & \multicolumn{2}{c}{$k = 1$} && \multicolumn{2}{c}{$k = 64$} \\  \cmidrule{2-3} \cmidrule{5-6}
                     & Time & Comm. && Time  & Comm. \\ \midrule
    \textbf{AlexNet} & 0.91 & 0.002 && 1.09  & 0.16 \\
    \textbf{VGG-16}  & 2.14 & 0.056  && 11.76 & 3.60 \\ \bottomrule
  \end{tabular}
  \caption{Running time (in seconds) and total communication (in GB) for batch private inference on CIFAR-10
  using a batch size of $k$.}
  \label{tab:amortized-cifar}
\end{table}

\begin{table}
  \centering
  \begin{tabular}{lcc c cc}
    \toprule 
    & \multicolumn{2}{c}{$k = 1$} && \multicolumn{2}{c}{$k = 8$} \\  \cmidrule{2-3} \cmidrule{5-6}
                         & Time & Comm. && Time & Comm. \\ \midrule
    \textbf{ResNet-50}   & 9.31 & 3.08  && 42.99 & 24.7 \\
    \textbf{ResNet-101}  & 17.62 & 4.64  && 72.99 & 37.2 \\ 
    \textbf{ResNet-152}  & 25.77 & 6.56 && 105.20 & 52.5 \\ \bottomrule
  \end{tabular}
  \caption{Running time (in seconds) and total communication (in GB) for batch private inference on ImageNet
  using a batch size of $k$.}
  \label{tab:amortized-imagenet}
\end{table}

\begin{table*}[h!]
  \begin{minipage}[t]{\linewidth}
  \renewcommand{\thempfootnote}{\fnsymbol{mpfootnote}}
  \renewcommand{\footnoterule}{}
  \centering
  \begin{tabular}{lcc c cc c cc c cc c cc}
    \toprule
    & \multicolumn{2}{c}{{\bf LeNet (MNIST)}} && \multicolumn{2}{c}{\bf AlexNet (CIFAR-10)} && \multicolumn{2}{c}{\bf VGG-16 (CIFAR-10)} && \multicolumn{2}{c}{\bf AlexNet (TI)} && 
      \multicolumn{2}{c}{\bf VGG-16 (TI)} \\
      \cmidrule{2-3} \cmidrule{5-6} \cmidrule{8-9} \cmidrule{11-12} \cmidrule{14-15}
    & Time & Comm. && Time       & Comm. && Time & Comm. && Time & Comm. && Time & Comm. \\ \midrule
    {\bf \textsc{Falcon}\footnote{The provided
    implementation of \textsc{Falcon} does not support computing the gradients for the
    output layer, so the \textsc{Falcon} measurements only include the time for computing the gradients
    for intermediate layers. All measurements for \textsc{Falcon} are taken {\em without} batch normalization.}}
    & 14.90 & 0.346 && 62.37 & 0.621 && 360.83\mpfootnotemark[2] & 1.78\mpfootnotemark[2] && 415.67 & 2.35 && 359.60\mpfootnotemark[3] & 1.78\mpfootnotemark[3] \\ 
    \rowcolor{yellow!40!white}
    {\bf \oursystem} & {2.21} & 1.14 && {2.91} & 1.37 && {12.14}\footnote{Using a smaller batch size of 32 images per iteration (due to GPU memory limitations). We make the same batch size adjustment for \textsc{Falcon}.}
    & 7.55\mpfootnotemark[2] && {11.30} & 6.98 &&
    {13.89}\footnote{Using a smaller batch size of 8 images per iteration (due to GPU memory limitations). We make the same batch size adjustment for \textsc{Falcon}.}
    & 7.59\mpfootnotemark[3] \\ \midrule
    {\bf Plaintext} & 0.0025 & --- && 0.0049 & --- && 0.0089 & --- && 0.0099 & --- && 0.0086 & --- \\ \bottomrule
  \end{tabular}
  \end{minipage}
  \caption{Running time (in seconds) and total communication (in GB)
  for a single iteration of private training with a batch size of 128 images
  for different models, datasets, and systems in a LAN setting. The ``TI'' dataset refers to the Tiny ImageNet
  dataset~\cite{TinyImageNet}. The plaintext measurements correspond
  to the cost of training on plaintext data on the GPU.
  Performance numbers for \textsc{Falcon} are
  obtained by running its reference implementation~\cite{FalconCode}
  on three compute-optimized AWS instances in a LAN environment (see \cref{sec:benchmarks-main}).
  As discussed in \cref{sec:exp-setup}, when testing the performance of \textsc{CryptGPU},
  we replace max pooling with average pooling in all of the networks. }
  \label{tab:training-comp}
\end{table*}

\paragraph{Private training.}
We expect GPUs to have a larger advantage in the setting of private
training (just like modern deep learning, training is much more challenging than inference and thus,
more reliant on hardware acceleration). We measure the time needed for a {\em single iteration} of private
backpropagation (\cref{sec:crypto-protocols-train})
on a batch size of 128 images 
for several dataset/model configurations and summarize our results in \cref{tab:training-comp}
(together with measurements for the equivalent plaintext protocol). 
We only compare with \textsc{Falcon} because \textsc{CrypTFlow} does not support private
training. We note that the public implementation of the \textsc{Falcon} system~\cite{FalconCode}
does {\em not} include support for computing the cross-entropy loss function for backpropagation. However,
given the gradients for the output layer, the provided implementation supports gradient
computation for {\em intermediate} layers.
Thus, our measurements for the \textsc{Falcon} system only includes the cost of
computing the gradients for intermediate layers and not for the output layer; this provides a {\em lower
bound} on the running time of using \textsc{Falcon} for private training. Our system supports the full
backpropagation training algorithm.

Our system achieves a considerable speedup over \textsc{Falcon} in multiple
settings, especially over larger models and datasets. For instance,
to train AlexNet on Tiny ImageNet, a single iteration of (private) backpropagation 
completes in 11.30s with \oursystem and 6.9 minutes using
\textsc{Falcon}. For context, privately training AlexNet on Tiny ImageNet
(100,000 examples) would just take over a week
($\approx 10$ days) using \oursystem while it would take
over a year ($\approx 375$ days) using \textsc{Falcon}
(assuming 100 epochs over the training set).

On the larger VGG-16 network,
our system is constrained by the amount of available GPU memory.
Our system currently supports a maximum batch size of 32 when training VGG-16
on \mbox{CIFAR-10} and a maximum 
batch size of 8 when training on Tiny ImageNet. To establish a fair comparison
when comparing our system against \textsc{Falcon}
for privately training VGG-16, we apply the same batch size adjustment.
As shown in \cref{tab:training-comp}, when training VGG-16,
our system is $30\times$ faster when training on CIFAR-10
and $26\times$ when training on Tiny ImageNet. 
Reducing the memory overhead of our protocol and augmenting
it with support for {\em multiple} GPUs (as is standard
for modern deep learning) will enable better scalability. We leave this as an interesting direction
for future work.

Like the setting of private inference, there still remains a large gap
(roughly $2000\times$) between the costs of private training and plaintext
training (on the GPU). Designing new cryptographic protocols that can take
even better advantage of GPU parallelism will be important for closing this
gap.

\paragraph{Private training breakdown.} In \cref{tab:falcon-comp-fine}, we
provide a fine-grained breakdown of the costs of processing the different
layers in a single iteration of private training. Not surprisingly, the
primary advantage of our GPU-based protocol compared to the CPU-based protocol
of \textsc{Falcon} is in the computation of the linear layers. In the settings
we consider, evaluation of the linear layers is between $25\times$ and
$70\times$ faster with our system. The linear layers are the {\em primary} bottleneck
in \textsc{Falcon}, and account for $86\%$ to $99\%$ of the overall computational cost.
In \oursystem, the computational costs are more evenly split between the linear
layers and the non-linear layers.

For the pooling layers, the performance difference between \oursystem and \textsc{Falcon}
can be partially attributed to the the fact that \textsc{Falcon} uses {\em max pooling}
rather than {\em average pooling}. As discussed in \cref{sec:exp-setup},
average pooling is a linear function and simpler to evaluate privately. However,
our measurements show that \oursystem maintains a (significant) performance edge
even if we exclude the cost of the pooling layers from the running time of \textsc{Falcon}.

Finally, for the ReLU layers, the CPU-based protocol in \textsc{Falcon}
compares very favorably with the ReLU protocol in \oursystem, and even outperforms
our protocol on the smaller models and datasets.
Having a ReLU protocol that can
better take advantage of GPU parallelism will likely improve the performance
of our protocol. As described in \cref{sec:crypto-protocols-inf}, our ReLU protocol
relies on an arithmetic-to-binary share conversion, which is less GPU-friendly compared
to bilinear operations. The ReLU protocol from \textsc{Falcon} relies on different
techniques and it is interesting whether their approach can be adapted to be efficiently
computed on the GPU.

\paragraph{Avenues for improvement.}
Compared to \textsc{Falcon}, our private training protocol is more communication-intensive. \textsc{Falcon}
develops a number of specialized cryptographic protocols to substantially reduce the 
communication in their protocols. We believe it is an interesting question to study whether the protocols
developed in \textsc{Falcon} are ``GPU-friendly'' and can benefit from GPU acceleration.

\oursystem does not currently support batch normalization during private training, so we
do not report private training benchmarks on the ResNet-family of models.\footnote{Note that we can still
perform private inference for a model that is trained using batch normalization. Namely, the normalization
parameters are secret-shared (as part of the model) and applying batch normalization just corresponds
to an affine transformation.} Developing a GPU-friendly protocol
for batch normalization is an interesting avenue for further work and an important step towards supporting
private training of the ResNet family of models.
We are not aware of any system that currently supports private training over ResNet.

\begin{table*}[t]
  \begin{minipage}[t]{\linewidth}
    \renewcommand{\thempfootnote}{\fnsymbol{mpfootnote}}
    \renewcommand{\footnoterule}{}
    \centering
  \begin{tabular}{lcc c cc c cc c cc}
    \toprule
    & \multicolumn{2}{c}{\bf Linear} && \multicolumn{2}{c}{\bf Pooling} && \multicolumn{2}{c}{\bf ReLU} && \multicolumn{2}{c}{\bf Softmax} \\
    \cmidrule{2-3} \cmidrule{5-6} \cmidrule{8-9} \cmidrule{11-12}  
    & \textsc{Falcon} & \oursystem && \textsc{Falcon} & \oursystem && \textsc{Falcon} & \oursystem && \textsc{Falcon} & \oursystem \\ \midrule
    {\bf LeNet (MNIST)}   & 13.07   & 0.49 && 1.34  & 0.076 && 0.47 & 1.00 && --- & 0.53\\
    {\bf AlexNet (CIFAR)} & 59.23   & 0.86 && 2.65  & 0.077 && 0.41 & 1.33 && --- & 0.55\\
    {\bf VGG-16 (CIFAR)}\footnote{Using a smaller batch size 32 images per iteration (due to GPU memory limitations). We make the same batch size adjustment for \textsc{Falcon}.}  & 355.16  & 6.33 && 2.86  & 0.21 && 5.40 & 4.74 && --- & 0.53\\
    {\bf AlexNet (TI)}    & 402.45 & 5.60 && 10.20 & 0.37 && 1.92 & 4.16 && --- & 1.04\\
    {\bf VGG-16 (TI)}\footnote{Using a smaller batch size 8 images per iteration (due to GPU memory limitations). We make the same batch size adjustment for \textsc{Falcon}.}     & 355.84  & 7.61 && 2.87  & 0.32 && 5.37 & 4.73 && --- & 0.98 \\ \bottomrule
  \end{tabular}
  \end{minipage}
  \caption{Runtime (in seconds) of \textsc{Falcon}~\cite{WTBKMR21} and \oursystem for evaluating
  the linear, pooling, ReLU, and softmax layers for different models and datasets during private {\em training}.
  The ``linear'' layers include the convolution and the fully-connected
  layers. The ``pooling'' layer refers to max pooling in \textsc{Falcon}, and average pooling in \oursystem.
  The implementation of \textsc{Falcon}~\cite{FalconCode} does not currently support softmax evaluation
  (and correspondingly, gradient computation for the output layer). Performance numbers for \textsc{Falcon} are
  obtained by running its reference implementation~\cite{FalconCode}
  on three compute-optimized AWS instances in a LAN environment (see \cref{sec:benchmarks-main}).
  }
  \label{tab:falcon-comp-fine}
\end{table*}

\subsection{Microbenchmarks}
\label{sec:micro}

To quantify the advantage of keeping all of the computation
on the GPU, we compare the running time of the
MPC protocols for evaluating convolutions
(i.e., the linear layers) and for evaluating ReLU (i.e., the primary non-linear layer)
on the CPU vs. the GPU. For convolutions, we study the effect of both the input dimension
as well as the batch size. We use the same experimental setup described in
\cref{sec:exp-setup}
for all of the experiments in this section.

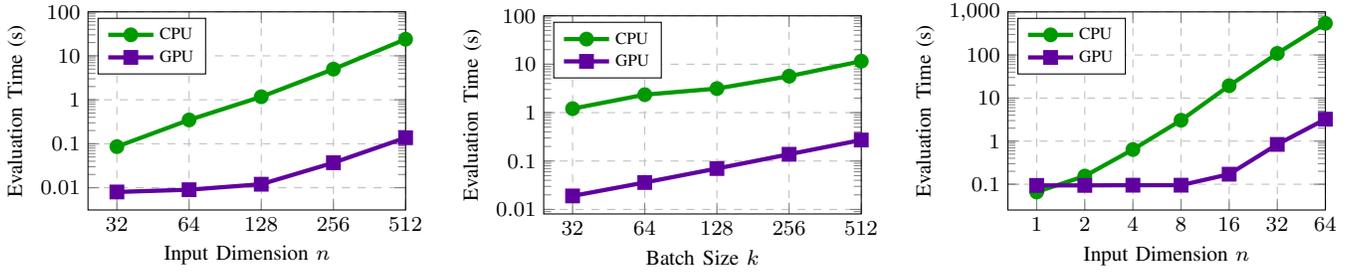
\begin{figure*}[h]
  \footnotesize
  \begin{subfigure}[b]{0.32\textwidth}
    \captionsetup{format=hang}
    \centering
    \begin{tikzpicture}
    \begin{semilogxaxis}[
        xlabel = {Input Dimension $n$},
        ylabel = {Evaluation Time (s)},
        xmax = 512,
        ymax = 100,
        ytick distance = 10,
        ymode = log, log ticks with fixed point,
        xtick = data,
        legend pos = north west,
        legend style = {nodes={scale=0.8, transform shape}},
        ymajorgrids = true,
        xmajorgrids = true,
        grid style = dashed,
        ylabel near ticks, 
        xlabel near ticks,
        width  = {\textwidth},
        height = {15em},
        legend cell align = {left},
    ]
     
    \addplot[plot1] table[x=dim,y=cpu] {\convsmalldata};
    \addplot[plot2] table[x=dim,y=gpu]{\convsmalldata};

    \legend{CPU, GPU}
     
    \end{semilogxaxis}
    \end{tikzpicture}
    \caption{Convolution on an $n \times n \times 3$ input with an
    $11 \times 11$ kernel, $64$ output channels, $4 \times 4$ stride,
    and $2 \times 2$ padding.}
    \label{fig:micro-cpu-gpu-convlow}
  \end{subfigure}
  \hspace{0.2em}
  \begin{subfigure}[b]{0.32\textwidth}
    \captionsetup{format=hang}
    \centering
    \begin{tikzpicture}
    \begin{semilogxaxis}[
        xlabel = {Batch Size $k$},
        ylabel = {Evaluation Time (s)},
        xmax = 512,
        ymax = 100,
        ytick distance = 10,
        ymode = log, log ticks with fixed point,
        xtick = data,
        legend pos = north west,
        legend style = {nodes={scale=0.8, transform shape}},
        ymajorgrids = true,
        xmajorgrids = true,
        grid style = dashed,
        ylabel near ticks,
        xlabel near ticks,
        width  = {\textwidth},
        height = {15em},
        legend cell align = {left},
    ]
     
    \addplot[plot1] table[x=batch,y=cpu] {\convbatchdata};
    \addplot[plot2] table[x=batch,y=gpu]{\convbatchdata};

    \legend{CPU, GPU}
     
    \end{semilogxaxis}
    \end{tikzpicture}
    \caption{Convolution on batch of $k$ $32 \times 32 \times 3$ inputs
    with an $11 \times 11$ kernel, $64$ output channels, $4 \times 4$ stride, and 
    $2 \times 2$ padding.}
    \label{fig:micro-cpu-gpu-convbatch}
\end{subfigure}
\hspace{0.2em}
\begin{subfigure}[b]{0.32\textwidth}
    \captionsetup{format=hang}
    \centering
    \begin{tikzpicture}
    \begin{semilogxaxis}[
        xlabel = {Input Dimension $n$},
        ylabel = {Evaluation Time (s)},
        title = {},
        xmax = 64,
        ymax = 1000,
        ytick distance = 10,
        ymode = log, log ticks with fixed point,
        xtick = data,
        legend pos = north west,
        legend style = {nodes={scale=0.8, transform shape}},
        ymajorgrids = true,
        xmajorgrids = true,
        grid style = dashed,
        ylabel near ticks,
        xlabel near ticks,
        width  = {\textwidth},
        height = {15em},
        legend cell align = {left},
    ]
     
    \addplot[plot1] table[x=dim,y=cpu] {\convbigdata};
    \addplot[plot2] table[x=dim,y=gpu]{\convbigdata};

    \legend{CPU, GPU}
     
    \end{semilogxaxis}
    \end{tikzpicture}
    \caption{Convolution on an $n \times n \times 512$ input with a $3 \times 3$ kernel,
    $512$ output channels, $1 \times 1$ stride, and $1 \times 1$ padding.}
    \label{fig:micro-cpu-gpu-convhigh}
  \end{subfigure}
  \caption{Comparison of total protocol execution time (in a LAN setting)
  for privately evaluating convolutions on the CPU and the GPU. Parameters for convolution
  kernels are chosen based on parameters in AlexNet~\cite{KSH12}. The stride and padding parameters
  specify how the filter is applied to the input. All of the figures are log-log plots.}
  \label{fig:micro-cpu-gpu}
\end{figure*}

\paragraph{Private convolution: GPU vs. CPU.}
For convolutions, we consider two types of convolutions: (1) convolutions with a large
receptive field (filter size) but a relatively small number of input/output channels;
and (2) convolutions with a small receptive field, but a large number of input/output channels.
Convolutions of the first type are generally used in the initial layers of the CNN while filters
of the second type are used in the later layers of the CNN. Note that when implementing convolutions
on the CPU, we do {\em not} break up the 64-bit secret-shared tensor into 16-bit
blocks (as we do in the GPU setting; see \cref{sec:sys-design}). We provide the microbenchmarks
in \cref{fig:micro-cpu-gpu}.

From \cref{fig:micro-cpu-gpu-convlow,fig:micro-cpu-gpu-convhigh}, we see that for small
inputs, the computational cost of the private convolution protocol is comparable on both the
GPU and the GPU. While there is only a $10\times$
speed-up for convolutions between a small $32 \times 32 \times 3$ input with a stack
of $64$ filters, the gap grows quickly as the input size
increases; for instance, increasing the input size to that of a Tiny ImageNet
instance $(64 \times 64 \times 3)$, the GPU-based protocol is nearly $40 \times$ faster.
Scaling to a $512 \times 512 \times 3$ image, the GPU-based protocol is $174\times$ faster
than the CPU-based protocol (from 23.9s on the CPU to 0.14s on the GPU).
An analogous trend holds when we consider convolutions with
a large number of input/output channels: for small inputs, the running times of the CPU- and
GPU-based protocols are quite comparable, but for large inputs (e.g., a $64 \times 64 \times 512$ input),
the GPU-based protocol is $168\times$ faster (from 543s on the CPU to just 3.2s on the GPU).

We additionally note that for small instances, the protocol running time on the GPU is essentially
constant---this is due to the parallelism. Only after the input becomes sufficiently large do we start
seeing increases in the running time based on the size of the input. In contrast, the CPU running time
always scales with the size of the input.

Similar speedups are present when we consider convolutions on batches of inputs (this is important
for training and for {\em batch} inference). For a fixed input size ($32 \times 32 \times 3$)
and kernel size ($11 \times 11$), we observe a $10\times$ speed-up for running the private
convolution protocol on a single input using the GPU, but a $40\times$ to $60\times$ speed-up
when we consider a batch of anywhere from 32 to 512 inputs. As an example, to evaluate a convolution
over a batch of 512 inputs with this set of parameters, we require 11.6s on the CPU and only
0.27s on the GPU. We refer to \cref{fig:micro-cpu-gpu-convbatch} for the full comparison.

\pgfplotstableread{
  dim gpu    cpu
  32  0.008  0.086
  64  0.009  0.348
  128 0.012  1.172
  256 0.037  4.986
  512 0.137  23.877
}\convsmalldata

\pgfplotstableread{
  dim gpu    cpu
  1   0.094  0.066
  2   0.094  0.155
  4   0.095  0.64
  8   0.095  3.055
  16  0.171  19.363  
  32  0.84   108.819  
  64  3.237  543.531  
}\convbigdata

\pgfplotstableread{
  batch gpu    cpu
  32    0.019  1.208
  64    0.036  2.351
  128   0.07   3.142
  256   0.138  5.657
  512   0.272  11.551
}\convbatchdata

\paragraph{Private ReLU: GPU vs. CPU.} Previous privacy-preserving ML
systems like $\textsc{Delphi}$~\cite{MLSZP20}
leveraged GPUs to accelerate convolutions, but still executed the non-linear
steps (e.g., ReLU computations) on the CPU. Here, we argue that with a carefully-chosen
set of cryptographic protocols, we can also take advantage of GPU parallelism
to accelerate the non-linear computations.
To illustrate this, we compare the running time of our private ReLU protocol
on the CPU vs. the GPU. As described in \cref{sec:crypto-protocols-inf}, private ReLU evaluation
of ReLU on a large block of neurons (e.g., output by the convolutional layer) corresponds
to evaluating a large number of point-wise Boolean operations on secret-shared binary tensors.
Such operations naturally benefit from GPU parallelism.

\begin{figure}[h]
  \centering
  \footnotesize
  \begin{tikzpicture}
  \begin{axis}[
      xlabel = {Input Size (millions of elements)},
      ylabel = {Evaluation Time (s)},
      xmin = 0, xmax = 32,
      ymin = 0, ymax = 150,
      ymode = log, log ticks with fixed point,
      xtick distance = 4,
      minor x tick num = 4,
      ytick distance = 10,
      legend pos = south east,
      ymajorgrids = true,
      xmajorgrids = true,
      grid style = dashed,
      ylabel near ticks,
      xlabel near ticks,
      width  = {0.5\textwidth},
      height = {15em},
  ]
   
  \addplot[plot1] table[x=size,y=cpu] {\relucomputedata};
  \addplot[plot2] table[x=size,y=gpu] {\relucomputedata};
  \legend{CPU, GPU}
   
  \end{axis}
  \end{tikzpicture}
  \caption{Comparison of total protocol execution time (in a LAN setting) on the CPU vs. the GPU
  for point-wise evaluation of the private ReLU protocol on different-sized inputs.}
  \label{fig:micro-relu}
\end{figure}
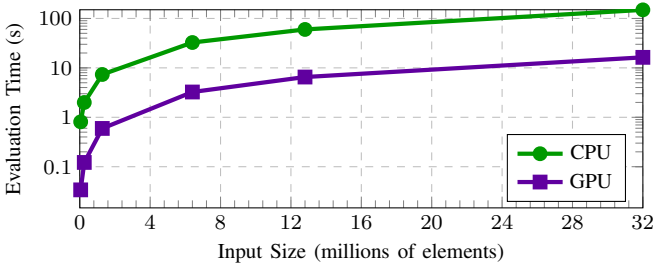

We measure the time it takes to privately-evaluate
ReLU on different numbers of secret-shared inputs (ranging from 50,000 to 32,000,000). The full results
are shown in \cref{fig:micro-relu}. For ReLU evaluation, we see a $16\times$ speedup when evaluating
ReLU on a block of 256,000 inputs (from 2s on the CPU to 0.12s on the GPU). As we scale up to a block
with 32~million inputs (250 MB of data), there is a $9\times$ speedup on the GPU, with the absolute
running time dropping from 149s on the CPU to just 16.3s on the GPU.

\pgfplotstableread{
  size      gpu    cpu
  0.0512    0.034  0.81
  0.256     0.122  1.995
  1.28      0.595  7.352
  6.4       3.249  32.597
  12.8      6.526  59.93
  32        16.332 149.183
}\relucomputedata

\iftoggle{fullversion}{

\iftoggle{fullversion}{
  \subsection{Accuracy of Privacy-Preserving Protocols}
}{
  \section{Accuracy of Privacy-Preserving Protocols}
}
\label{sec:acc}

Several of the underlying protocols in \oursystem are not exact and can
introduce a small amount of error: using fixed-point encodings to approximate
floating-point arithmetic, the share-truncation protocol from $\aby$, and the
approximation to the softmax function. While
we have chosen our parameters (e.g., the fixed-point precision) to reduce the
likelihood of errors, we validate our parameter choices with an empirical
analysis. In the following, we will often measure the difference between an
output $z_{\text{priv}}$ computed using \oursystem with the output
$z_{\text{plain}}$ of a plaintext version of the same computation
(using 64-bit floating-point values).
We define the {\em absolute error} between $z_{\text{priv}}$ and $z_{\text{plain}}$
as $\abs{z_{\text{priv}} - z_{\text{plain}}}$ and the {\em relative error}
to be $\abs{z_{\text{priv}} - z_{\text{plain}}} / z_{\text{plain}}$.

\paragraph{Fixed point precision.} As discussed at the beginning of \cref{sec:exp},
\oursystem emulates floating-point
computations by encoding values using a fixed-point representation
using $t = 20$ bits of fractional precision.
The fixed-point computations over the integers are embedded
into operations on secret-shared values
over the ring $\ring$. The modulus $n$ must be large enough to support multiplication
(and more generally, convolution and matrix multiplication) of plaintext values
without triggering a modular reduction.
In \oursystem, $n = 2^{64}$, so shares are represented by 64-bit integers.

Previous privacy-preserving protocols like \textsc{Falcon}~\cite{WTBKMR21}
and \textsc{Delphi}~\cite{MLSZP20}
use a smaller number of bits of fixed-point precision (e.g., 13 bits and 15 bits, respectively).
In turn, they are able to work with arithmetic shares over a 32-bit ring as opposed
to a 64-bit ring. This reduces communication (since shares are half as large) and in our
model, also saves computation (recall from \cref{sec:sys-design} that we need to split
up tensors of 64-bit integers into 4 tensors of 16-bit integers in order to use
existing CUDA kernels for deep learning).

Using fewer number of bits of precision reduces
the accuracy of the protocol outputs, especially when scaling to deep architectures and
large inputs. To analyze the effect the number of bits of fixed-point precision $t$ has on
the accuracy of the outputs of our system (i.e., the values of the output layer),
we compute the average relative error between the output values output
by \oursystem to those computed using 
the plaintext inference protocol on a small example (AlexNet over CIFAR-10)
as well as a large example (ResNet-50 on ImageNet). Our results are summarized
in \cref{fig:fp-precision}. 

\cref{fig:fp-precision} shows that for a relatively shallow model like
AlexNet on the CIFAR-10 dataset, it is sufficient to use 12 to 14 bits of fixed-point
precision (e.g., the parameter setting in~\cite{WTBKMR21}).
The relative error in this case between the outputs computed by the private inference protocol
and the plaintext computation is around 1\%. However, when we scale up to a model
like ResNet-50 on ImageNet, the average relative error in the model outputs
increases $5 \times$ to almost 5\%. We further remark that we are only measuring the relative
error in a single forward pass over the network (inference). Larger errors
would be expected in the case of private training when the protocol needs to run
multiple forward and backward passes. In this work, we use $t = 20$ bits of fixed-point
precision which ensures that the average relative error for private inference
over ResNet-50 on ImageNet is under 0.02\%. Our analysis indicates that scaling up
to deeper architectures and operating over larger datasets will require a greater number
of bits of precision in the underlying fixed-point representation. For instance,
to keep the average relative error under
1\% for ResNet-50 on ImageNet, we require at least 15 bits of fixed-point
precision. As such, to prevent overflows in the arithmetic evaluation over secret-shared
data for deep networks, a 32-bit ring is no longer sufficient.

\pgfplotstableread{
  precision alexnet    resnet 
  10        6.01       18.37
  12        1.36       4.90
  14        0.37       1.20
  16        0.07       0.31
  18        0.02       0.07
  20        0.005      0.02
}\precisiondata

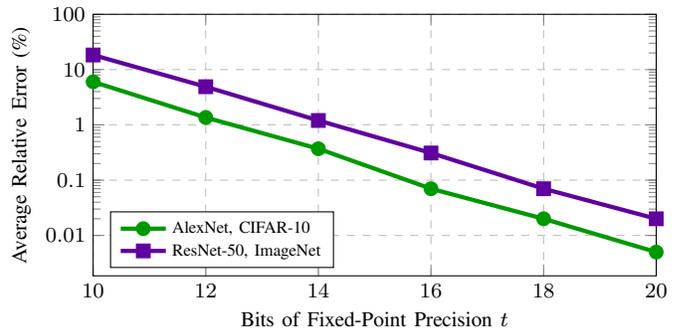
\begin{figure}[t]
  \footnotesize
  \centering
  \begin{tikzpicture}
  \begin{axis}[
      xlabel = {Bits of Fixed-Point Precision $t$},
      ylabel = {Average Relative Error (\%)},
      xmin = 10,
      xmax = 20,
      ymin = 0,
      ymax = 100,
      ytick distance = 10,
      ymode = log, log ticks with fixed point,
      xtick = data,
      legend pos = south west,
      legend style = {nodes={scale=0.8, transform shape}},
      ymajorgrids = true,
      xmajorgrids = true,
      grid style = dashed,
      ylabel near ticks, 
      xlabel near ticks,
      width  = {0.5 \textwidth},
      height = {18em},
      legend cell align = {left}
  ]
   
  \addplot[plot1] table[x=precision,y=alexnet] {\precisiondata};
  \addplot[plot2] table[x=precision,y=resnet]{\precisiondata};

  \legend{{AlexNet, CIFAR-10}, {ResNet-50, ImageNet}}
   
  \end{axis}
  \end{tikzpicture}
  \caption{Average relative error between the
  model outputs computed using the private inference protocol
  in \oursystem with $t$ bits of fixed-point precision (i.e., an integer $x \in \R$ is represented
  as the nearest integer to $x \cdot 2^t$)
  and the output computed using {\em plaintext}
  floating-point inference. Analysis based on evaluating AlexNet on CIFAR-10 and 
  ResNet-50 on ImageNet, and averaged over 10 randomly-chosen instances. }
  \label{fig:fp-precision}
  \iftoggle{fullversion}{}{\vspace{-1em}}
\end{figure}

\paragraph{Privacy-preserving inference.} To evaluate the accuracy of our
private inference protocol, we compare the average relative error between the 
outputs of our private inference protocol using
ResNet-50, ResNet-101, and ResNet-152 on ImageNet and compare those against the values
obtained from plaintext evaluation. We additionally compute the accuracy of the predictions
(using the standard metrics of Top-1 and Top-5 accuracy---i.e., the model succeeds if the actual
class of an example coincides with the most likely class predicted by the model or among the
top 5 most likely classes predicted by the model). The results are summarized in \cref{tab:accuracy}.
In particular, for our chosen set of parameters, we observe that the average relative error in the classifier
output is at most 0.021\%, and in all cases we tested (100 randomly-chosen images from the ImageNet test set),
both the Top-1 accuracy and the Top-5 accuracy {\em exactly} match that of the plaintext model.

\begin{table}
  \centering
  \begin{tabular}{lccc}
    \toprule 
      & {\bf ResNet-50} & {\bf ResNet-101} & {\bf ResNet-152} \\ \midrule
    \textbf{Average Relative Error} &  0.015\% & 0.020\%  & 0.021\% \\ \midrule
    \textbf{Top-1 Acc. (\oursystem)} & 78\%  & 82\% & 79\% \\ 
    \textbf{Top-1 Acc. (Plaintext)}  & 78\%  & 82\% & 79\% \\ \midrule
    \textbf{Top-5 Acc. (\oursystem)} & 92\%  & 90\% & 93\% \\
    \textbf{Top-5 Acc. (Plaintext)}  & 92\%  & 90\% & 93\% \\ \bottomrule
  \end{tabular}
  \caption{Comparison of outputs of \oursystem's 
  private inference protocol on ImageNet with the ResNet models with those of the plaintext
  algorithm (using \pytorch).
  The average relative error is computed between the outputs of the private
  inference protocol and those of the plaintext execution (on the same input).
  The Top-1 and Top-5 accuracies for both settings
  are computed based on the outputs of model inference with respect
  to the ground truth label. The measurements are taken over
  a random set of 100 examples drawn from the ImageNet validation set.}
  \label{tab:accuracy}
  \iftoggle{fullversion}{}{\vspace{-1em}}
\end{table}

\pgfplotsset{%
  trainpriv/.style = {%
    green!60!black,
    mark=none,
    very thick,
    line join=round,
  },
  trainplain/.style = {%
    purple!50!blue,
    mark=none,
    very thick,
    line join=round
  }
}

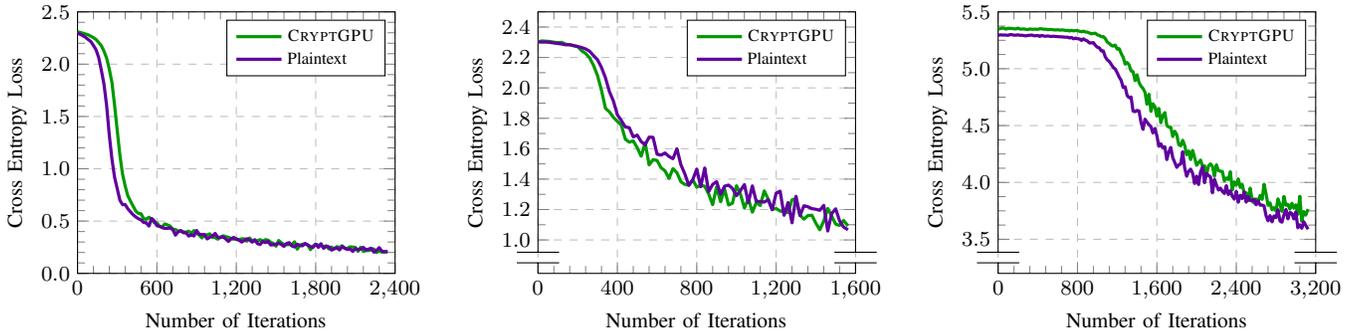
\begin{figure*}
  \begin{subfigure}[b]{0.32\textwidth}
  \footnotesize
  \centering
  \begin{tikzpicture}
  \begin{axis}[
      xlabel = {Number of Iterations},
      ylabel = {Cross Entropy Loss},
      xmin = 0,
      xmax = 2400,
      xtick distance = 600,
      minor x tick num = 4,
      ymin = 0,
      ymax = 2.5,
      ytick distance = 0.5,
      minor y tick num = 4,
      y tick label style = {/pgf/number format/.cd, fixed, fixed zerofill, precision=1, /tikz/.cd},
      legend pos = north east,
      legend style = {nodes={scale=0.8, transform shape}},
      ymajorgrids = true,
      xmajorgrids = true,
      grid style = dashed,
      ylabel near ticks, 
      xlabel near ticks,
      width  = {\textwidth},
      height = {18em},
      legend cell align = {left}
  ]
   
  \addplot[trainpriv] table[x=iter,y=priv] {\mnisttraining};
  \addplot[trainplain] table[x=iter,y=plain]{\mnisttraining};

  \legend{\oursystem, Plaintext}
  \end{axis}
  \end{tikzpicture}
  \caption{LeNet on MNIST (trained for 5 epochs with a batch size of 128).}
  \end{subfigure}
  \hspace{0.2em}
  \begin{subfigure}[b]{0.32\textwidth}
  \footnotesize
  \centering
  \begin{tikzpicture}
  \begin{axis}[
      xlabel = {Number of Iterations},
      ylabel = {Cross Entropy Loss},
      xmin = 0,
      xmax = 1600,
      xtick distance = 400,
      minor x tick num = 4,
      ymin = 0.78,
      ymax = 2.5,
      ytick distance = 0.2,
      axis y discontinuity=parallel,
      ytickmin = 1,
      minor y tick num = 1,
      y tick label style = {/pgf/number format/.cd, fixed, fixed zerofill, precision=1, /tikz/.cd},
      legend pos = north east,
      legend style = {nodes={scale=0.8, transform shape}},
      ymajorgrids = true,
      xmajorgrids = true,
      grid style = dashed,
      ylabel near ticks, 
      xlabel near ticks,
      width  = {\textwidth},
      height = {18em},
      legend cell align = {left}
  ]
   
  \addplot[trainpriv] table[x=iter,y=priv] {\cifartraining};
  \addplot[trainplain] table[x=iter,y=plain]{\cifartraining};

  \legend{\oursystem, Plaintext}
  \end{axis}
  \end{tikzpicture}
  \caption{AlexNet on CIFAR-10 (trained for 1 epoch with a batch size of 32).}
  \end{subfigure}
  \hspace{0.2em}
  \begin{subfigure}[b]{0.32\textwidth}
  \footnotesize
  \centering
  \begin{tikzpicture}
  \begin{axis}[
      xlabel = {Number of Iterations},
      ylabel = {Cross Entropy Loss},
      xmin = 0,
      xmax = 3200,
      xtick distance = 800,
      minor x tick num = 4,
      ymin = 3.2,
      ymax = 5.5,
      ytick distance = 0.5,
      axis y discontinuity=parallel,
      ytickmin = 3.5,
      minor y tick num = 3,
      y tick label style = {/pgf/number format/.cd, fixed, fixed zerofill, precision=1, /tikz/.cd},
      legend pos = north east,
      legend style = {nodes={scale=0.8, transform shape}},
      ymajorgrids = true,
      xmajorgrids = true,
      grid style = dashed,
      ylabel near ticks, 
      xlabel near ticks,
      width  = {\textwidth},
      height = {18em},
      legend cell align = {left}
  ]
   
  \addplot[trainpriv] table[x=iter,y=priv] {\titraining};
  \addplot[trainplain] table[x=iter,y=plain]{\titraining};

  \legend{\oursystem, Plaintext}
  \end{axis}
  \end{tikzpicture}
  \caption{AlexNet on Tiny ImageNet (trained for 1 epoch on a batch size of 32).}
  \end{subfigure}
  \caption{Moving average of the 
  cross-entropy loss as a function of the number of training
  iterations using \oursystem and using
  a plaintext protocol for different models and datasets.
  In each setting, we use the same initialization and learning rate (for stochastic gradient descent) for both private and plaintext training. For LeNet, we use a random initialization.
  For the AlexNet experiments, we use \pytorch's default AlexNet architecture~\cite{PyTorchAlexNet}
  for both the plaintext training
  and the private training experiments (which is a variant of the standard
  AlexNet architecture described in~\cite{KSH12}). We use \pytorch's pre-trained
  initialization for AlexNet as our initialization.
  The moving average is computed over a window
  of size 20 (i.e., the value reported for iteration $i$ is
  the average of the cross entropy loss on iterations $i - 10, \ldots, i + 9$).}
  \label{fig:training-loss}
  \iftoggle{fullversion}{}{\vspace{-1em}}
\end{figure*}

\begin{table}[h!]
  \begin{minipage}[t]{\linewidth}
  \renewcommand{\thempfootnote}{\fnsymbol{mpfootnote}}
  \renewcommand{\footnoterule}{}
  \centering
  \begin{tabular}{lccc}
    \toprule
      & {\bf Baseline} & {\bf \oursystem} & {\bf Plaintext} \\ \midrule
    {\bf LeNet, MNIST}\footnote{Trained for 5 epochs (2345 iterations) with a batch size of 128.}
      & 10\% & 93.97\% & 93.34\% \\ \midrule
    {\bf AlexNet, CIFAR-10}\footnote{Trained for 1 epoch (1563 iterations) with a batch size of 32.}
      & 10\% & 59.60\% & 59.77\% \\
    {\bf AlexNet, Tiny ImageNet}\footnote{Trained for 1 epoch (3125 iterations) with a batch size of 32.} 
      & 2\% & 17.82\% & 17.51\% \\ \bottomrule
  \end{tabular}
  \end{minipage}
  \caption{Validation set accuracy for different models trained using \oursystem
  and the plaintext training algorithm. For each configuration, both training
  approaches use the same initialization and learning rate
  (for stochastic gradient descent). For LeNet, we use a random initialization.
  For the AlexNet experiments, we use \pytorch's default AlexNet architecture~\cite{PyTorchAlexNet}
  for both the plaintext training
  and the private training experiments. Here, we use \pytorch's pre-trained
  weights for AlexNet to speed up convergence. We also report the baseline
  accuracy for each configuration (i.e., accuracy of the ``random-guess'' algorithm). Note that
  training for more iterations will increase the accuracy; the intent of this comparison
  is to demonstrate a close similarity in model accuracies for the the model output by the private
  training protocol with the model output by plaintext training
  after a few thousand iterations of stochastic gradient descent.}
  \label{tab:training-acc}
  \iftoggle{fullversion}{}{\vspace{-1em}}
\end{table}

\paragraph{Privacy-preserving training.} We perform a similar set of
experiments to evaluate the accuracy of our private training protocol.
In \cref{fig:training-loss}, we plot the value of the cross-entropy
loss function for a model trained using the private training protocol
of \oursystem as well as for a model trained using the plaintext
training algorithm (using the same initialization and
learning rate for the underlying stochastic gradient descent
optimizer). \cref{fig:training-loss} shows that the value of the loss
function is slightly higher initially for private training, but the
overall progression closely follows that of plaintext training.

In addition to comparing the evolution of the loss function, we also compare
the model accuracies (as measured on the validation set) for the models
trained using \oursystem and using the plaintext training algorithm (again
with same initialization and learning rate as above). Our results are
summarized in \cref{tab:training-acc}. On all of the models/datasets we
considered, the accuracy of the model output by
\oursystem closely matches that of the plaintext evaluation.
These experiments indicate that \oursystem efficiently and accurately
supports {\em end-to-end} private
training for models like AlexNet over moderately-large datasets like Tiny ImageNet.

\paragraph{Average pooling vs. max pooling.}
As discussed in \cref{sec:exp-setup}, we use average pooling in place of max
pooling in the models we consider. To evaluate whether the choice of pooling
makes a significant difference on model performance, we use \pytorch to train
the AlexNet and VGG-16 networks over the CIFAR-10 dataset where we replace all
of the max pooling layers with average pooling layers. The resulting model
accuracy on the CIFAR-10 test set is shown in \cref{tab:pooling}. In
particular, we observed a $3\%$ drop in accuracy (from 76\% to 73\%) for
AlexNet and a $1\%$ increase in accuracy with VGG-16 (from 82\% to 83\%). This
indicates that using average pooling in place of max pooling does not lead to
a significant degradation of model performance. We note also that in contrast to
AlexNet and VGG-16 which use max pooling exclusively, the more recent ResNets
use average pooling in all but the initial layer.

\begin{table}
  \centering
    \begin{tabular}{lcc}
    \toprule
    & {\bf Max Pooling} & {\bf Average Pooling} \\ \midrule
    {\bf AlexNet} & 76.15\% & 73.35\% \\
    {\bf VGG-16}   & 82.37\% & 83.17\% \\ \bottomrule
  \end{tabular}
  \caption{Validation set accuracy for plaintext training of AlexNet
  and VGG-16 over the CIFAR-10 dataset using max pooling vs. average pooling.
  All networks were trained using
  50 epochs using a standard stochastic gradient descent (SGD) optimizer in \pytorch.}
  \label{tab:pooling}
  \iftoggle{fullversion}{}{\vspace{-3em}}
\end{table}
}{
  \paragraph{Accuracy analysis.} In \cref{sec:acc}, we provide additional experiments
  to measure the accuracy of our privacy-preserving machine learning protocols.
}


\section{Related Work}
\label{sec:related}

Privacy-preserving machine learning is a special case of secure computation and
can be solved via general cryptographic approaches such as secure 2-party computation (2PC)~\cite{Yao86},
secure multiparty computation~\cite{GMW87,BGW88} or fully homomorphic encryption~\cite{Gen09}.
While powerful, these general approaches incur significant overhead, and much of the work in
developing concretely-efficient protocols for scalable privacy-preserving machine learning
have focused on more specialized approaches (that still rely on the general building blocks for
designing sub-protocols). We survey some of these techniques here.

\paragraph{Privacy-preserving inference.} Many recent works have developed specific protocols
for the problem of private inference for deep learning models~(c.f.,~\cite{GDLLNW16,MZ17,LJLA17,CGRST17,MR18,RWTSSK18,JVC18,WGC19,CCPS19,RSCLLK19,MLSZP20,KRCGRS20,DEK20,PS20,BCPS20,CRS20,KVHSIM20,BCLMJ20,WTBKMR21} and the references therein). These works operate in a variety of different models and architectures:
some works consider a 2-party setting (e.g., \cite{MZ17,RWTSSK18,JVC18,MLSZP20}), others consider a 3-party (e.g., \cite{MR18,WGC19,WTBKMR21,KRCGRS20,CCPS19,PS20}) or a 4-party setting (e.g.,~\cite{BCPS20,CRS20}). Some frameworks
assume that the model is held in the clear (e.g., \cite{JVC18,MLSZP20})
while others (including this work) support secret-shared models (e.g., \cite{WTBKMR21,KRCGRS20}).
With the recent exceptions of \textsc{Falcon}~\cite{WTBKMR21} and \textsc{CrypTFlow}~\cite{KRCGRS20}, these
existing approaches only consider privacy-preserving inference using shallow neural networks (e.g., less than 10 layers)
on relatively small datasets (at the scale of MNIST~\cite{MNIST} or CIFAR~\cite{CIFAR}).
Our focus in this work is designing
privacy-preserving machine learning protocols that are able to support inference over modern deep learning models
(which typically contain {\em tens of millions of parameters} and over a hundred layers) on large datasets
(i.e., at the scale of ImageNet~\cite{ImageNet}, one of the de facto standards for state-of-the art computer vision).
As shown in \cref{sec:benchmarks-main}, our system
outperforms both \textsc{Falcon} and \textsc{CrypTFlow} for inference over sufficiently-large models and datasets.

\paragraph{Privacy-preserving training.} Compared to private inference, privacy-preserving training of
deep neural networks is a considerably more challenging and computationally-intensive 
problem and has received comparably less attention. 
Of the aforementioned works, only a few~\cite{MZ17,MR18,CCPS19,WGC19,BCPS20,PS20,CRS20,WTBKMR21} support
privacy-preserving training. Among these systems, the only one that scales beyond MNIST/CIFAR is
\textsc{Falcon}~\cite{WTBKMR21}, which is the first system (to our knowledge) that supports privacy-preserving training
at the scale of (Tiny) ImageNet and for models as large as AlexNet~\cite{KSH12} and VGG-16~\cite{SZ14}.
Our work is the first framework
to leverage GPUs to demonstrate significantly better scalability to privately train deep networks
over large datasets.

\paragraph{Privacy-preserving machine learning using GPUs.}
Most of the works on privacy-preserving machine learning are CPU-based and do not leverage GPU
acceleration. We discuss some notable exceptions. Some works~\cite{BVMA18,BCLMJ20} use GPUs to 
accelerate homomorphic evaluation of convolutional neural networks on MNIST. \textsc{Delphi}~\cite{MLSZP20} uses
GPUs to compute linear layers (i.e., convolutions) to support private inference; however, they still
perform non-linear operations (e.g., ReLU evaluation) on the CPU and moreover, their scheme assumes
the model to be public (and only the input is hidden). Our design philosophy in this work is to keep
{\em all} of the computations on the GPU through a careful choice of ``GPU-friendly'' cryptographic
protocols. Slalom~\cite{TB19} shows how to integrate a trusted computing base (e.g., Intel SGX) with
GPUs to enable fast private inference of neural networks (by offloading convolutions to the GPU and performing
non-linear operations within the trusted enclave). Recent works proposing scalable private training and inference
protocols highlight the use of GPUs as an important way for further scalability~\cite{WTBKMR21,KRCGRS20}.
Our system is the first to support private training and inference {\em entirely} on the GPU.

\paragraph{Model stealing and inversion attacks.} We note that MPC protocols 
can only hide the inputs to the computation (e.g., the model or the dataset) up to
what can be inferred from the output. Several recent works~\cite{AMSVVF15,FJR15,TZJRR16,RRK18,JCBKP19}
have shown how black-box access
to a model (in the case of an private inference service) can allow an adversary to learn
information about the model or even recover its training data. Differentially-private
training algorithms~\cite{SS15,ACGMMTZ16} provide one defense against certain types of these attacks.
Our focus in this work is on
protecting the {\em computation} itself and ensure that there is no {\em additional} leakage
about the inputs other than through the output. It is an interesting question to design a private
training/inference protocol that also provides robustness against specific
classes of model stealing/inversion attacks.


\section{Conclusion}
In this paper, we introduce \oursystem, a new MPC framework that implements {\em all} of the cryptographic operations (both linear {\em and} non-linear) on the GPU. \oursystem is built on top of PyTorch~\cite{PyTorch19} and \crypten~\cite{KVHSIM20} to make it easy to use for machine learning developers and researchers. Our experiments show that leveraging GPUs can significantly accelerate the private training and inference for modern deep learning and make it practical to run privacy-preserving deep learning at the scale of ImageNet and with complex networks. In addition, our systematic analysis of different cryptographic protocols provides new insights for designing ``GPU-friendly'' cryptographic protocols for deep learning. This will be an important step towards bridging the roughly $1000\times$ 
gap that still remains between private machine learning
and plaintext machine learning (on the GPU).

\section*{Acknowledgments}
We thank Pavel Belevich, Shubho Sengupta, and Laurens van der Maaten
for their feedback on system design and providing helpful pointers.
D.~J.~Wu is supported by NSF CNS-1917414.

\bibliographystyle{ieeetr}
\bibliography{paper}

\begin{thebibliography}{10}

\bibitem{MZ17}
P.~Mohassel and Y.~Zhang, ``{SecureML}: {A} system for scalable
  privacy-preserving machine learning,'' in {\em {IEEE} Symposium on Security
  and Privacy}, pp.~19--38, 2017.

\bibitem{MR18}
P.~Mohassel and P.~Rindal, ``{ABY}\({}^{\mbox{3}}\): {A} mixed protocol
  framework for machine learning,'' in {\em {ACM} {CCS}}, pp.~35--52, 2018.

\bibitem{WGC19}
S.~Wagh, D.~Gupta, and N.~Chandran, ``{SecureNN}: 3-party secure computation
  for neural network training,'' {\em Proc. Priv. Enhancing Technol.},
  vol.~2019, no.~3, pp.~26--49, 2019.

\bibitem{MLSZP20}
P.~Mishra, R.~Lehmkuhl, A.~Srinivasan, W.~Zheng, and R.~A. Popa, ``Delphi: {A}
  cryptographic inference service for neural networks,'' in {\em {USENIX}
  Security}, pp.~2505--2522, 2020.

\bibitem{KRCGRS20}
N.~Kumar, M.~Rathee, N.~Chandran, D.~Gupta, A.~Rastogi, and R.~Sharma,
  ``{CrypTFlow}: Secure tensorflow inference,'' in {\em {IEEE} Symposium on
  Security and Privacy}, pp.~336--353, 2020.

\bibitem{WTBKMR21}
S.~Wagh, S.~Tople, F.~Benhamouda, E.~Kushilevitz, P.~Mittal, and T.~Rabin,
  ``{FALCON:} honest-majority maliciously secure framework for private deep
  learning,'' {\em Proc. Priv. Enhancing Technol.}, vol.~2021, 2021.

\bibitem{GMW87}
O.~Goldreich, S.~Micali, and A.~Wigderson, ``How to play any mental game or {A}
  completeness theorem for protocols with honest majority,'' in {\em {STOC}},
  pp.~218--229, 1987.

\bibitem{BGW88}
M.~Ben{-}Or, S.~Goldwasser, and A.~Wigderson, ``Completeness theorems for
  non-cryptographic fault-tolerant distributed computation (extended
  abstract),'' in {\em {STOC}}, pp.~1--10, 1988.

\bibitem{SS15}
R.~Shokri and V.~Shmatikov, ``Privacy-preserving deep learning,'' in {\em {ACM}
  {CCS}}, pp.~1310--1321, 2015.

\bibitem{ACGMMTZ16}
M.~Abadi, A.~Chu, I.~J. Goodfellow, H.~B. McMahan, I.~Mironov, K.~Talwar, and
  L.~Zhang, ``Deep learning with differential privacy,'' in {\em {ACM} {CCS}},
  pp.~308--318, 2016.

\bibitem{MNIST}
Y.~{LeCun}, C.~Cortes, and C.~J. Burges, ``The {MNIST} database.''
  \url{http://yann.lecun.com/exdb/mnist/}.

\bibitem{CIFAR}
A.~Krizhevsky, ``Learning multiple layers of features from tiny images,'' 2009.

\bibitem{ImageNet}
O.~Russakovsky, J.~Deng, H.~Su, J.~Krause, S.~Satheesh, S.~Ma, Z.~Huang,
  A.~Karpathy, A.~Khosla, M.~S. Bernstein, A.~C. Berg, and F.~Li, ``Imagenet
  large scale visual recognition challenge,'' {\em Int. J. Comput. Vis.},
  vol.~115, no.~3, pp.~211--252, 2015.

\bibitem{HZRS16}
K.~He, X.~Zhang, S.~Ren, and J.~Sun, ``Deep residual learning for image
  recognition,'' in {\em {CVPR}}, pp.~770--778, 2016.

\bibitem{TinyImageNet}
F.-F. Li, A.~Karpathy, and J.~Johnson, ``Tiny {ImageNet} visual recognition
  challenge,'' 2017.

\bibitem{KSH12}
A.~Krizhevsky, I.~Sutskever, and G.~E. Hinton, ``Imagenet classification with
  deep convolutional neural networks,'' in {\em {NeurIPS}}, pp.~1106--1114,
  2012.

\bibitem{SZ14}
K.~Simonyan and A.~Zisserman, ``Very deep convolutional networks for
  large-scale image recognition,'' in {\em {ICLR}}, 2015.

\bibitem{LBDHHHJ89}
Y.~LeCun, B.~E. Boser, J.~S. Denker, D.~Henderson, R.~E. Howard, W.~E. Hubbard,
  and L.~D. Jackel, ``Backpropagation applied to handwritten zip code
  recognition,'' {\em Neural Comput.}, vol.~1, no.~4, pp.~541--551, 1989.

\bibitem{CPS06}
K.~Chellapilla, S.~Puri, and P.~Simard, ``High performance convolutional neural
  networks for document processing,'' 2006.

\bibitem{CMGS10}
D.~C. Ciresan, U.~Meier, L.~M. Gambardella, and J.~Schmidhuber, ``Deep, big,
  simple neural nets for handwritten digit recognition,'' {\em Neural Comput.},
  vol.~22, no.~12, pp.~3207--3220, 2010.

\bibitem{PyTorch19}
A.~Paszke, S.~Gross, F.~Massa, A.~Lerer, J.~Bradbury, G.~Chanan, T.~Killeen,
  Z.~Lin, N.~Gimelshein, L.~Antiga, A.~Desmaison, A.~K{\"{o}}pf, E.~Yang,
  Z.~DeVito, M.~Raison, A.~Tejani, S.~Chilamkurthy, B.~Steiner, L.~Fang,
  J.~Bai, and S.~Chintala, ``{PyTorch}: An imperative style, high-performance
  deep learning library,'' in {\em {NeurIPS}}, pp.~8024--8035, 2019.

\bibitem{TensorFlow16}
M.~Abadi, A.~Agarwal, P.~Barham, E.~Brevdo, Z.~Chen, C.~Citro, G.~S. Corrado,
  A.~Davis, J.~Dean, M.~Devin, S.~Ghemawat, I.~J. Goodfellow, A.~Harp,
  G.~Irving, M.~Isard, Y.~Jia, R.~J{\'{o}}zefowicz, L.~Kaiser, M.~Kudlur,
  J.~Levenberg, D.~Man{\'{e}}, R.~Monga, S.~Moore, D.~G. Murray, C.~Olah,
  M.~Schuster, J.~Shlens, B.~Steiner, I.~Sutskever, K.~Talwar, P.~A. Tucker,
  V.~Vanhoucke, V.~Vasudevan, F.~B. Vi{\'{e}}gas, O.~Vinyals, P.~Warden,
  M.~Wattenberg, M.~Wicke, Y.~Yu, and X.~Zheng, ``Tensorflow: Large-scale
  machine learning on heterogeneous distributed systems,'' {\em CoRR},
  vol.~abs/1603.04467, 2016.

\bibitem{TPU}
``Cloud tensor processing units (tpus).''
  \url{https://cloud.google.com/tpu/docs/tpus}.

\bibitem{KVHSIM20}
B.~Knott, S.~Venkataraman, A.~Hannun, S.~Sengupta, M.~Ibrahim, and L.~van~der
  Maaten, ``{CrypTen}: Secure multi-party computation meets machine learning,''
  in {\em Proceedings of the {NeurIPS} Workshop on Privacy-Preserving Machine
  Learning}, 2020.

\bibitem{ISN89}
M.~Ito, A.~Saito, and T.~Nishizeki, ``Secret sharing scheme realizing general
  access structure,'' {\em Electronics and Communications in Japan (Part III:
  Fundamental Electronic Science)}, vol.~72, no.~9, pp.~56--64, 1989.

\bibitem{AFLNO16}
T.~Araki, J.~Furukawa, Y.~Lindell, A.~Nof, and K.~Ohara, ``High-throughput
  semi-honest secure three-party computation with an honest majority,'' in {\em
  {ACM} {CCS}}, pp.~805--817, 2016.

\bibitem{CUDA}
``{CUDA} libraries documentation.''
  \url{https://docs.nvidia.com/cuda-libraries/index.html}.

\bibitem{Yao86}
A.~C. Yao, ``How to generate and exchange secrets (extended abstract),'' in
  {\em {FOCS}}, pp.~162--167, 1986.

\bibitem{CCPS19}
H.~Chaudhari, A.~Choudhury, A.~Patra, and A.~Suresh, ``{ASTRA:} high throughput
  3pc over rings with application to secure prediction,'' in {\em {ACM} {CCS}},
  pp.~81--92, 2019.

\bibitem{PS20}
A.~Patra and A.~Suresh, ``{BLAZE:} blazing fast privacy-preserving machine
  learning,'' in {\em {NDSS}}, 2020.

\bibitem{KMR11}
S.~Kamara, P.~Mohassel, and M.~Raykova, ``Outsourcing multi-party
  computation,'' {\em {IACR} Cryptol. ePrint Arch.}, vol.~2011, p.~272, 2011.

\bibitem{cuBLAS}
``{cuBLAS}.'' \url{https://docs.nvidia.com/cuda/cublas/index.html}.

\bibitem{cuDNN}
``{cuDNN}.''
  \url{https://docs.nvidia.com/deeplearning/cudnn/developer-guide/index.html}.

\bibitem{DSZ15}
D.~Demmler, T.~Schneider, and M.~Zohner, ``{ABY} - {A} framework for efficient
  mixed-protocol secure two-party computation,'' in {\em {NDSS}}, 2015.

\bibitem{Bea91}
D.~Beaver, ``Efficient multiparty protocols using circuit randomization,'' in
  {\em {CRYPTO}}, pp.~420--432, 1991.

\bibitem{TB19}
F.~Tram{\`{e}}r and D.~Boneh, ``Slalom: Fast, verifiable and private execution
  of neural networks in trusted hardware,'' in {\em {ICLR}}, 2019.

\bibitem{FLNW17}
J.~Furukawa, Y.~Lindell, A.~Nof, and O.~Weinstein, ``High-throughput secure
  three-party computation for malicious adversaries and an honest majority,''
  in {\em {EUROCRYPT}}, pp.~225--255, 2017.

\bibitem{Can00}
R.~Canetti, ``Security and composition of multiparty cryptographic protocols,''
  {\em J. Cryptol.}, vol.~13, no.~1, pp.~143--202, 2000.

\bibitem{Gol04}
O.~Goldreich, {\em The Foundations of Cryptography - Volume 2: Basic
  Applications}.
\newblock Cambridge University Press, 2004.

\bibitem{NH10}
V.~Nair and G.~E. Hinton, ``Rectified linear units improve restricted boltzmann
  machines,'' in {\em {ICML}}, pp.~807--814, 2010.

\bibitem{GBC16}
I.~Goodfellow, Y.~Bengio, and A.~Courville, {\em Deep Learning}.
\newblock MIT Press, 2016.
\newblock \url{http://www.deeplearningbook.org}.

\bibitem{PyTorchAES}
``{PyTorch/CSPRNG}.'' \url{https://github.com/pytorch/csprng}.

\bibitem{SMDS11}
J.~K. Salmon, M.~A. Moraes, R.~O. Dror, and D.~E. Shaw, ``Parallel random
  numbers: as easy as 1, 2, 3,'' in {\em Conference on High Performance
  Computing Networking, Storage and Analysis, {SC}}, pp.~16:1--16:12, 2011.

\bibitem{LBBH98}
Y.~{LeCun}, L.~Bottou, Y.~Bengio, and P.~Haffner, ``Gradient-based learning
  applied to document recognition,'' {\em Proceedings of the IEEE}, vol.~86,
  no.~11, pp.~2278--2324, 1998.

\bibitem{FalconCode}
{Sameer Wagh and Shruti Tople and Fabrice Benhamouda and Eyal Kushilevitz and
  Prateek Mittal and Tal Rabin}, ``Falcon: Honest-majority maliciously secure
  framework for private deep learning.''
\newblock Available at \url{https://github.com/snwagh/falcon-public}.

\bibitem{PyTorchAlexNet}
P.~Team, ``Alexnet.'' \url{https://pytorch.org/hub/pytorch_vision_alexnet/}.

\bibitem{Gen09}
C.~Gentry, {\em A fully homomorphic encryption scheme}.
\newblock PhD thesis, Stanford University, 2009.
\newblock \url{crypto.stanford.edu/craig}.

\bibitem{GDLLNW16}
R.~Gilad{-}Bachrach, N.~Dowlin, K.~Laine, K.~E. Lauter, M.~Naehrig, and
  J.~Wernsing, ``Cryptonets: Applying neural networks to encrypted data with
  high throughput and accuracy,'' in {\em {ICML}}, pp.~201--210, 2016.

\bibitem{LJLA17}
J.~Liu, M.~Juuti, Y.~Lu, and N.~Asokan, ``Oblivious neural network predictions
  via minionn transformations,'' in {\em {ACM} {CCS}}, pp.~619--631, 2017.

\bibitem{CGRST17}
N.~Chandran, D.~Gupta, A.~Rastogi, R.~Sharma, and S.~Tripathi, ``Ezpc:
  Programmable, efficient, and scalable secure two-party computation for
  machine learning.'' Cryptology ePrint Archive, Report 2017/1109, 2017.
\newblock \url{https://eprint.iacr.org/2017/1109}.

\bibitem{RWTSSK18}
M.~S. Riazi, C.~Weinert, O.~Tkachenko, E.~M. Songhori, T.~Schneider, and
  F.~Koushanfar, ``Chameleon: {A} hybrid secure computation framework for
  machine learning applications,'' in {\em {ACM} {CCS}}, pp.~707--721, 2018.

\bibitem{JVC18}
C.~Juvekar, V.~Vaikuntanathan, and A.~Chandrakasan, ``{GAZELLE:} {A} low
  latency framework for secure neural network inference,'' in {\em {USENIX}
  Security Symposium}, pp.~1651--1669, 2018.

\bibitem{RSCLLK19}
M.~S. Riazi, M.~Samragh, H.~Chen, K.~Laine, K.~E. Lauter, and F.~Koushanfar,
  ``{XONN:} xnor-based oblivious deep neural network inference,'' in {\em
  {USENIX} Security Symposium}, pp.~1501--1518, 2019.

\bibitem{DEK20}
A.~P.~K. Dalskov, D.~Escudero, and M.~Keller, ``Secure evaluation of quantized
  neural networks,'' {\em Proc. Priv. Enhancing Technol.}, vol.~2020, no.~4,
  pp.~355--375, 2020.

\bibitem{BCPS20}
M.~Byali, H.~Chaudhari, A.~Patra, and A.~Suresh, ``{FLASH:} fast and robust
  framework for privacy-preserving machine learning,'' {\em Proc. Priv.
  Enhancing Technol.}, vol.~2020, no.~2, pp.~459--480, 2020.

\bibitem{CRS20}
H.~Chaudhari, R.~Rachuri, and A.~Suresh, ``Trident: Efficient 4pc framework for
  privacy preserving machine learning,'' in {\em {NDSS}}, 2020.

\bibitem{BCLMJ20}
A.~A. Badawi, J.~Chao, J.~Lin, C.~F. Mun, S.~J. Jie, B.~H.~M. Tan, X.~Nan,
  A.~M.~M. Khin, and V.~Chandrasekhar, ``Towards the alexnet moment for
  homomorphic encryption: {HCNN}, the first homomorphic cnn on encrypted data
  with gpus,'' {\em IEEE Transactions on Emerging Topics in Computing}, 2020.

\bibitem{BVMA18}
A.~A. Badawi, B.~Veeravalli, C.~F. Mun, and K.~M.~M. Aung, ``High-performance
  {FV} somewhat homomorphic encryption on gpus: An implementation using
  {CUDA},'' {\em {IACR} Trans. Cryptogr. Hardw. Embed. Syst.}, vol.~2018,
  no.~2, pp.~70--95, 2018.

\bibitem{AMSVVF15}
G.~Ateniese, L.~V. Mancini, A.~Spognardi, A.~Villani, D.~Vitali, and G.~Felici,
  ``Hacking smart machines with smarter ones: How to extract meaningful data
  from machine learning classifiers,'' {\em Int. J. Secur. Networks}, vol.~10,
  no.~3, pp.~137--150, 2015.

\bibitem{FJR15}
M.~Fredrikson, S.~Jha, and T.~Ristenpart, ``Model inversion attacks that
  exploit confidence information and basic countermeasures,'' in {\em {ACM}
  {CCS}}, pp.~1322--1333, 2015.

\bibitem{TZJRR16}
F.~Tram{\`{e}}r, F.~Zhang, A.~Juels, M.~K. Reiter, and T.~Ristenpart,
  ``Stealing machine learning models via prediction apis,'' in {\em {USENIX}
  Security Symposium}, pp.~601--618, 2016.

\bibitem{RRK18}
B.~D. Rouhani, M.~S. Riazi, and F.~Koushanfar, ``{DeepSecure}: Scalable
  provably-secure deep learning,'' in {\em Annual Design Automation
  Conference}, pp.~1--6, 2018.

\bibitem{JCBKP19}
M.~Jagielski, N.~Carlini, D.~Berthelot, A.~Kurakin, and N.~Papernot,
  ``High-fidelity extraction of neural network models,'' {\em CoRR},
  vol.~abs/1909.01838, 2019.

\end{thebibliography}

\appendices
\crefalias{section}{appendix}
\iftoggle{fullversion}{}{}
\iftoggle{fullversion}{}{}
\iftoggle{fullversion}{}{}
\iftoggle{fullversion}{

\section{Network Architecture}
\label{sec:network-arch}

As discussed in \cref{sec:exp-setup}, some of the models we consider (e.g., AlexNet and VGG-16)
were designed for ImageNet, and are not directly compatible with smaller datasets such as
CIFAR-10 and Tiny ImageNet. As such, when training or running inference with these models
on the smaller datasets, we make adjustments to their ``head architecture'' (i.e., the fully-connected
classification layers at the top of the network).
In all settings, we keep the same ``base architecture'' (adapted from their
description in the original papers~\cite{KSH12,SZ14}).
We describe the base AlexNet architecture we use in \cref{fig:alexnet-base} and
the head architectures for the different datasets in \cref{fig:alexnet-head-arch}.
We describe the base VGG-16 architecture we use in
\cref{fig:vgg-base} and the head architectures for the different
datasets in \cref{fig:vgg-head-arch}.

\begin{figure*}[b]
    \centering
    \begin{tabular}{lclc}
        \toprule
        {\bf Layer} & {\bf Input Dimension} & \bf{Description} & \bf{Output Dimension} \\ \midrule
        {\bf Convolution} & $32 \times 32 \times 3$ & $11\times 11$ kernel,
            $9 \times 9$ padding, $4 \times 4$ stride & $10 \times 10 \times 96$  \\
        \textbf{ReLU} & $10 \times 10 \times 96$ & $\relu(\cdot)$ on each input & $10 \times 10 \times 96$  \\
        \textbf{Average Pooling} & $10 \times 10 \times 96$  & $3 \times 3$ kernel, $2 \times 2$ stride & $4 \times 4 \times 96$  \\ \midrule
        \textbf{Convolution} & $4 \times 4 \times 96$ & $5 \times 5$ kernel, $1 \times 1$ padding, $1 \times 1$ stride & $2 \times 2 \times 256$  \\
        \textbf{ReLU} & $2 \times 2 \times 256$ & $\relu(\cdot)$ on each input & $2 \times 2 \times 256$ \\
        \textbf{Average Pooling} & $2 \times 2 \times 256$ & $2 \times 2$ kernel, $1 \times 1$ stride & $1 \times 1 \times 256$\\ \midrule
        \textbf{Convolution} & $1 \times 1 \times 256$ & $3 \times 3$ kernel, $1 \times 1$ padding, $1 \times 1$ stride & $1 \times 1 \times 384$ \\
        \textbf{ReLU} & $1 \times 1 \times 384$ & $\relu(\cdot)$ on each input & $1 \times 1 \times 384$\\ \midrule
        \textbf{Convolution} & $1 \times 1 \times 384$ & $3\times 3$ kernel, $1 \times 1$ padding, $1 \times 1$ stride & $1 \times 1 \times 384$ \\
        \textbf{ReLU} & $1 \times 1 \times 384$ & $\relu(\cdot)$ on each input & $1 \times 1 \times 384$ \\ \midrule
        \textbf{Convolution} & $1 \times 1 \times 384$ & $3\times 3$ kernel, $1 \times 1$ padding, $1 \times 1$ stride & $1 \times 1 \times 256$\\
        \textbf{ReLU} & $1 \times 1 \times 256$ & $\relu(\cdot)$ on each input & $1 \times 1 \times 256$\\ \bottomrule
    \end{tabular}
    \caption{AlexNet~\cite{KSH12} base architecture on CIFAR-10.
    The same architecture is also used
    for Tiny ImageNet and ImageNet, but applied
    to different input dimensions ($64 \times 64 \times 3$ for Tiny ImageNet
    and $224 \times 224 \times 3$ for ImageNet).
    The head architectures (classification layers) for CIFAR-10, Tiny ImageNet, and ImageNet
    vary (as a function of the input size and number of output classes) and are shown in
    \cref{fig:alexnet-head-arch}.}
    \label{fig:alexnet-base}
\end{figure*}

\begin{figure*}[b]
    \begin{subfigure}[b]{\textwidth}
        \centering
        \begin{tabular}{l c l c}
            \toprule
            {\bf Layer} & {\bf Input Dimension} & \bf{Description} & \bf{Output Dimension} \\ \midrule
            \textbf{Flatten} & $1 \times 1 \times 256$ & Flatten input into a single dimension & 256 \\ \midrule
            \textbf{Fully Connected} &  256 & $256\times 256$ matrix multiplication & 256 \\ 
            \textbf{ReLU} &  256 & $\relu(\cdot)$ on each input & 256 \\ \midrule
            \textbf{Fully Connected} & 256 & $256\times 256$ matrix multiplication & 256 \\ 
            \textbf{ReLU} &  256 & $\relu(\cdot)$ on each input & 256 \\ \midrule
            \textbf{Fully Connected} & 256 & $256\times 10$ matrix multiplication & 10 \\ \bottomrule
        \end{tabular}
        \caption{Head architecture for CIFAR-10}
        \label{fig:alexnet-cifar}
    \end{subfigure}

    \bigskip

    \begin{subfigure}[b]{\textwidth}
        \centering
        \begin{tabular}{l c l c}
            \toprule
            {\bf Layer} & {\bf Input Dimension} & \bf{Description} & \bf{Output Dimension} \\ \midrule
            \textbf{Average Pooling} & $4 \times 4 \times 256$ & $2 \times 2$ kernel, $2\times 2$ stride & $2 \times 2 \times 256$ \\
            \textbf{Flatten} & $2 \times 2 \times 256$ & Flatten input into a single dimension & 1024 \\\midrule
            \textbf{Fully Connected} &  1024 & $1024\times 1024$ matrix multiplication & 1024 \\ 
            \textbf{ReLU} &  1024 & $\relu(\cdot)$ on each input & 1024 \\ \midrule
            \textbf{Fully Connected} & 1024 & $1024\times 1024$ matrix multiplication & 1024 \\
            \textbf{ReLU} &  1024 & $\relu(\cdot)$ on each input & 1024 \\ \midrule
            \textbf{Fully Connected} & 1024 & $1024\times 200$ matrix multiplication & 200 \\ \bottomrule
        \end{tabular}
        \caption{Head architecture for Tiny ImageNet}
        \label{fig:alexnet-tinyin}
    \end{subfigure}

    \bigskip

    \begin{subfigure}[b]{\textwidth}
        \centering
        \begin{tabular}{l c l c}
            \toprule
            {\bf Layer} & {\bf Input Dimension} & \bf{Description} & \bf{Output Dimension} \\ \midrule
            \textbf{Average Pooling} & $24 \times 24 \times 256$ & $4 \times 4$ kernel, $4\times 4$ stride & $6 \times 6 \times 256$ \\
            \textbf{Flatten} & $6 \times 6 \times 256$ & Flatten input into a single dimension & 9216 \\\midrule
            \textbf{Fully Connected} &  9216 & $9216 \times 4096$ matrix multiplication & 4096 \\
            \textbf{ReLU} &  4096 & $\relu(\cdot)$ on each input & 4096 \\ \midrule
            \textbf{Fully Connected} & 4096 & $4096\times 4096$ matrix multiplication & 4096 \\
            \textbf{ReLU} &  4096 & $\relu(\cdot)$ on each input & 4096 \\ \midrule
            \textbf{Fully Connected} & 4096 & $4096\times 1000$ matrix multiplication & 1000 \\ \bottomrule
        \end{tabular}
        \caption{Head architecture for ImageNet}
        \label{fig:alexnet-imagenet}
    \end{subfigure}
    \caption{Head architecture of AlexNet for CIFAR-10, Tiny ImageNet, and ImageNet.}
    \label{fig:alexnet-head-arch}
\end{figure*}

\begin{figure*}[b]
    \centering
    \begin{tabular}{l c l c}
        \toprule
        {\bf Layer} & {\bf Input Dimension} & \bf{Description} & \bf{Output Dimension} \\ \midrule
        \textbf{Convolution} & $32 \times 32 \times 3$ & $3\times 3$ kernel, $1 \times 1$ padding, $1 \times 1$ stride & $32 \times 32 \times 64$\\ 
        \textbf{ReLU} & $32 \times 32 \times 64$ & $\relu(\cdot)$ on each input & $32 \times 32 \times 64$ \\ \midrule
        \textbf{Convolution} & $32 \times 32 \times 64$ & $3\times 3$ kernel, $1 \times 1$ padding, $1 \times 1$ stride & $32 \times 32 \times 64$\\ 
        \textbf{ReLU} &  $32 \times 32 \times 64$ & $\relu(\cdot)$ on each input &  $32 \times 32 \times 64$ \\ 
        \textbf{Average Pooling} &  $32 \times 32 \times 64$ & $2 \times 2$ kernel, $2\times 2$ stride &  $16 \times 16 \times 64$ \\ \midrule 
        \textbf{Convolution} & $16 \times 16 \times 64$ & $3\times 3$ kernel, $1 \times 1$ padding, $1 \times 1$ stride & $16 \times 16 \times 128$\\ 
        \textbf{ReLU} & $16 \times 16 \times 128$ & $\relu(\cdot)$ on each input & $16 \times 16 \times 128$ \\ \midrule
        \textbf{Convolution} & $16 \times 16 \times 128$ & $3\times 3$ kernel, $1 \times 1$ padding, $1 \times 1$ stride & $16 \times 16 \times 128$\\ 
        \textbf{ReLU} & $16 \times 16 \times 128$ & $\relu(\cdot)$ on each input & $16 \times 16 \times 128$\\ 
        \textbf{Average Pooling} & $16 \times 16 \times 128$ & $2 \times 2$ kernel, $2\times 2$ stride & $8 \times 8 \times 128$  \\ \midrule
        \textbf{Convolution} & $8 \times 8 \times 128$ & $3\times 3$ kernel, $1 \times 1$ padding, $1 \times 1$ stride & $8 \times 8 \times 256$\\
        \textbf{ReLU} & $8 \times 8 \times 256$ & $\relu(\cdot)$ on each input & $8 \times 8 \times 256$ \\ \midrule
        \textbf{Convolution} & $8 \times 8 \times 256$ & $3\times 3$ kernel, $1 \times 1$ padding, $1 \times 1$ stride & $8 \times 8 \times 256$\\ 
        \textbf{ReLU} & $8 \times 8 \times 256$ & $\relu(\cdot)$ on each input & $8 \times 8 \times 256$ \\ \midrule
        \textbf{Convolution} & $8 \times 8 \times 256$ & $3\times 3$ kernel, $1 \times 1$ padding, $1 \times 1$ stride & $8 \times 8 \times 256$\\ 
        \textbf{ReLU} & $8 \times 8 \times 256$ & $\relu(\cdot)$ on each input & $8 \times 8 \times 256$ \\ 
        \textbf{Average Pooling} & $8 \times 8 \times 256$ & $2 \times 2$ kernel, $2\times 2$ stride & $4 \times 4 \times 256$  \\ \midrule 
        \textbf{Convolution} & $4 \times 4 \times 256$ & $3\times 3$ kernel, $1 \times 1$ padding, $1 \times 1$ stride & $4 \times 4 \times 512$\\ 
        \textbf{ReLU} & $4 \times 4 \times 512$ &  $\relu(\cdot)$ on each input & $4 \times 4 \times 512$\\ \midrule
        \textbf{Convolution} & $4 \times 4 \times 512$ & $3\times 3$ kernel, $1 \times 1$ padding, $1 \times 1$ stride & $4 \times 4 \times 512$\\ 
        \textbf{ReLU} & $4 \times 4 \times 512$ &  $\relu(\cdot)$ on each input & $4 \times 4 \times 512$\\ \midrule
        \textbf{Convolution} & $4 \times 4 \times 512$ & $3\times 3$ kernel, $1 \times 1$ padding, $1 \times 1$ stride & $4 \times 4 \times 512$\\ 
        \textbf{ReLU} & $4 \times 4 \times 512$ &  $\relu(\cdot)$ on each input & $4 \times 4 \times 512$\\ 
        \textbf{Average Pooling} & $4 \times 4 \times 512$ & $2 \times 2$ kernel, $2\times 2$ stride & $2 \times 2 \times 512$  \\ \midrule 
        \textbf{Convolution} & $2 \times 2 \times 512$ & $3\times 3$ kernel, $1 \times 1$ padding, $1 \times 1$ stride & $2 \times 2 \times 512$\\ 
        \textbf{ReLU} & $2 \times 2 \times 512$ & $\relu(\cdot)$ on each input & $2 \times 2 \times 512$ \\ \midrule
        \textbf{Convolution} & $2 \times 2 \times 512$ & $3\times 3$ kernel, $1 \times 1$ padding, $1 \times 1$ stride & $2 \times 2 \times 512$\\ 
        \textbf{ReLU} & $2 \times 2 \times 512$ & $\relu(\cdot)$ on each input & $2 \times 2 \times 512$ \\ \midrule
        \textbf{Convolution} & $2 \times 2 \times 512$ & $3\times 3$ kernel, $1 \times 1$ padding, $1 \times 1$ stride & $2 \times 2 \times 512$\\ 
        \textbf{ReLU} & $2 \times 2 \times 512$ & $\relu(\cdot)$ on each input & $2 \times 2 \times 512$ \\ 
        \textbf{Average Pooling} & $2 \times 2 \times 512$ & $2 \times 2$ kernel, $2\times 2$ stride & $1 \times 1 \times 512$  \\ \bottomrule
    \end{tabular}
    \caption{VGG-16~\cite{SZ14} base architecture for CIFAR-10 inputs. The same architecture is also
    used for Tiny ImageNet and ImageNet, but applied to different input dimensions
    ($64 \times 64 \times 3$ for Tiny ImageNet and $224 \times 224 \times 3$ for ImageNet). The
    head architectures (classification layers) for CIFAR-10, Tiny ImageNet, and ImageNet
    vary (as a function of the input size and number of output classes) and are shown in \cref{fig:vgg-head-arch}.}
    \label{fig:vgg-base}
\end{figure*}

\begin{figure*}[b]
    \begin{subfigure}[b]{\textwidth}
    \centering
    \begin{tabular}{l c l c}
        \toprule
        {\bf Layer} & {\bf Input Dimension} & \bf{Description} & \bf{Output Dimension} \\ \midrule
        \textbf{Flatten} & $1 \times 1 \times 512$ & Flatten input into a single dimension & 512 \\\midrule
        \textbf{Fully Connected} &  512 & $512 \times 256$ matrix multiplication & 256 \\ 
        \textbf{ReLU} &  256 & $\relu(\cdot)$ on each input & 256 \\ \midrule
        \textbf{Fully Connected} & 256 & $256\times 256$ matrix multiplication & 256 \\ 
        \textbf{ReLU} &  256 & $\relu(\cdot)$ on each input & 256 \\ \midrule
        \textbf{Fully Connected} & 256 & $256\times 10$ matrix multiplication & 10 \\ \bottomrule
    \end{tabular}
    \caption{Head architecture for CIFAR-10.}
    \label{fig:vgg-cifar}
    \end{subfigure}

    \bigskip
    
    \begin{subfigure}[b]{\textwidth}
    \centering
    \begin{tabular}{l c l c}
        \toprule
        {\bf Layer} & {\bf Input Dimension} & \bf{Description} & \bf{Output Dimension} \\ \midrule
        \textbf{Average Pooling} & $2 \times 2 \times 512$ & $2 \times 2$ kernel, $2\times 2$ stride & $1 \times 1 \times 512$  \\  
        \textbf{Flatten} & $1 \times 1 \times 512$ & Flatten input into a single dimension & 512 \\\midrule
        \textbf{Fully Connected} &  512 & $512 \times 512$ matrix multiplication & 512 \\ 
        \textbf{ReLU} &  512 & $\relu(\cdot)$ on each input & 512 \\ \midrule
        \textbf{Fully Connected} & 512 & $512\times 512$ matrix multiplication & 512 \\ 
        \textbf{ReLU} &  512 & $\relu(\cdot)$ on each input & 512 \\ \midrule
        \textbf{Fully Connected} & 512 & $512\times 200$ matrix multiplication & 200 \\ \bottomrule
    \end{tabular}
    \caption{Head architecture for Tiny ImageNet.}
    \label{fig:vgg-tinyin}
    \end{subfigure}

    \bigskip
    
    \begin{subfigure}[b]{\textwidth}
    \centering
    \begin{tabular}{l c l c}
        \toprule
        {\bf Layer} & {\bf Input Dimension} & \bf{Description} & \bf{Output Dimension} \\ \midrule
        \textbf{Average Pooling} & $6 \times 6 \times 512$ & $2 \times 2$ kernel, $2\times 2$ stride & $3 \times 3 \times 512$  \\ 
        \textbf{Flatten} & $3 \times 3 \times 512$ & Flatten input into a single dimension & 4608 \\\midrule
        \textbf{Fully Connected} &  4608 & $4608 \times 4096$ matrix & 4096 \\ 
        \textbf{ReLU} &  4096 & $\relu(\cdot)$ on each input & 4096 \\ \midrule
        \textbf{Fully Connected} & 4096 & $4096\times 4096$ matrix & 4096 \\ 
        \textbf{ReLU} &  4096 & $\relu(\cdot)$ on each input & 4096 \\ \midrule
        \textbf{Fully Connected} & 4096 & $4096\times 1000$ matrix & 1000 \\ \bottomrule
    \end{tabular}
    \caption{Head architecture for ImageNet.}
    \label{fig:vgg-imagenet}
    \end{subfigure}

    \caption{Head architecture of VGG-16 for CIFAR-10, Tiny ImageNet, and ImageNet.}
    \label{fig:vgg-head-arch}
\end{figure*}

}{}

\end{document}